\documentclass[journal ]{new-aiaa}
\usepackage[utf8]{inputenc}
\usepackage{textcomp}

\usepackage{amsmath}
\usepackage{bm}
\usepackage{derivative}
\usepackage{siunitx}

\usepackage{tikz}
\usepackage{caption}
\usepackage{subcaption}
\usepackage{multirow}

\newcommand\revised[1]{\textcolor{black}{#1}}

\usepackage{graphicx}
\usepackage[version=4]{mhchem}
\usepackage{siunitx}
\usepackage{longtable,tabularx}
\title{Compressibility correction to the $k-\omega$ turbulence model that considers the wall-cooling effect}

\author{Zifei Yin\footnote{Associate Professor, School of Aeronautics and Astronautics, email: yinzifei@sjtu.edu.cn}}
\affil{Shanghai Jiao Tong University, Shanghai, 200240, China}

\begin{document}

\maketitle

\begin{abstract}
In supersonic and hypersonic flows, the near-wall density variation due to wall cooling poses a challenge for accurately predicting the near-wall velocity and temperature profiles using classical eddy viscosity turbulence models.
Compressible turbulent boundary layers are known to follow the universal wall law via semi-local transformation.
However, developing a turbulence model that predicts a velocity profile, which, via semi-local transformation, follows the universal wall law, remains challenging.
The current paper builds upon Danis-Durbin's practice of modifying the $\omega$ equation and proposes a simple modification to the $k-\omega$ two-equation model.
The formulation of the proposed modification involves dimensional analysis and the proper selection of the local length scale.
The newly introduced modification is used to modify the slope of the velocity profile starting from the viscous layer to above. 
It recovers a semi-local scaling of turbulent kinetic energy, viscosity, and eddy frequency, 
then achieves a very decent correction of the velocity profile in compressible turbulent channel flows, satisfying the universal wall law after applying Trettel \& Larsson's transformation.
The proposed new $k-\omega$ model can also improve the velocity and temperature predictions in strongly wall-cooled zero-pressure-gradient hypersonic turbulent boundary layers, compared to the original $k-\omega$ model.
Validation using the favorable and adverse pressure gradient boundary layers suggests that the model does not impose a negative effect on the original $k-\omega$ model.
\end{abstract}

\section*{Nomenclature}

{\renewcommand\arraystretch{1.0}
\noindent\begin{longtable*}{@{}l @{\quad=\quad} l@{}}
$\kappa$  & Von K\'{a}rm\'{a}n constant \\
$d$       & wall distance \\
$k$       & turbulent kinetic energy \\
$\omega$  & turbulent specific dissipation rate \\
VD        & Van Driest transformation \\
TL        & Trettel-Larsson transformation \\
$Y_{SL}$  & Trettel-Larsson's wall distance transformation, $Y_{SL}^{+}=\frac{{\rho}\left(\tau_w / {\rho}\right)^{1 / 2} y}{{\mu}}$\\
$Y_{TL}$  & Trettel-Larsson's velosity transformation, $U_{TL}^{+}=\int_0^{u^{+}}\left(\frac{{\rho}}{\rho_w}\right)^{1 / 2}\left[1+\frac{1}{2} \frac{1}{{\rho}} \frac{d {\rho}}{d y} y-\frac{1}{{\mu}} \frac{d {\mu}}{d y} y\right] d u^{+}$\\
$\mu$     & molecular dynamic viscosity \\
$\mu_t$   & turbulent dynamic viscosity \\
$\nu$     & molecular kinematic viscosity \\
$\nu_t$   & turbulent kinematic viscosity \\
$\alpha$  & power exponent \\
$Re_\tau$ & friction Reynolds number \\
$Re_\tau^*$ & friction Reynolds number after Trettel-Larsson transformation \\
$T_w$     & wall temperature \\
$T_c$     & channel centerline temperature \\
$T_\infty$ & freestream temperature \\
$\delta$  & boundary layer thickness \\
$Ma_\infty$ & freestream Mach number \\
$Ma_b$.   & bulk Mach number \\
$U_\infty$ & freestream velocity \\
$\rho_\infty$ & freestream density \\
$x_i$     & spatial coordinate \\
$Pr$      & prandtl number \\
$Pr_t$    & turbulent Prandtl number \\
$p$       & pressure \\
$u_i$     & velocity vector \\
$\rho$    & density \\
$T$       & temperature\\
$tau_{ij}$ & stress tensor \\
$q_j$     & heat flux \\
$C_p$     & specific heat at constant pressure \\
$C_f$     & skin friction coefficient\\
$C_h$     & wall heat transfer coefficient, $C_h=q_w /\left(\rho_{\infty} c_p U_{\infty}\left(T_r-T_w\right)\right)$\\
\end{longtable*}}

\section{Introduction}
In supersonic and especially hypersonic flows, the viscous heating of the boundary layer creates a high-temperature layer near the wall, which demands cooling techniques for the safety of the flight vehicle.
The design and optimization of thermal protection on hypersonic vehicle surfaces require prediction methods for the velocity and temperature profiles within the turbulent boundary layer.
The Reynolds-averaged Navier-Stokes (RANS) equations have been the workhorse in the past decades for engineering predictions of aerodynamic performance in the aerospace industry \citep{durbin2018some}.
Eddy-resolving simulations, including DNS, LES, and hybrid RANS/LES methods, are still costly for high Reynolds number flows \citep{menter2021overview}.
The RANS methods are expected to continue playing a crucial role in predicting supersonic and hypersonic engineering flows.

Viewing the physical foundation of the RANS methods, all the classical eddy viscosity models, such as the Prandtl's mixing length model, the Spalart-Allmaras model \cite{spalart1992one}, the $k-\omega$ model \citep{wilcox1998turbulence}, and Menter's shear stress transport model \citep{menter2003ten}, all incorporate the universal wall law in either an explicit or implicit manner.
However, although the universal wall law of turbulence has been very successful in incompressible flows, whether it can be directly applied to compressible flows is doubtful.
One of the reasons is that the velocity profiles of compressible wall turbulence, especially at high Mach numbers or with significant wall cooling effects, cannot be directly collapsed into the universal wall law. 
Only in recent years have promising efforts been made to collapse velocity and temperature profiles for compressible wall turbulence.
Another reason is that, in the development of classical turbulence models, compressibility effects are usually not directly incorporated into the model formulation.
Thus, some models face catastrophic failure when applied to flows where the near-wall compressibility effect is strong.
For some weakly compressible flows and even transonic flows, some turbulence models may still yield acceptable results and are therefore still used in certain engineering applications.
As a result, there is interest in developing modifications to classical turbulence models for supersonic and hypersonic flows, especially when the compressibility effect, or wall-cooling caused density gradient, is significant near the wall.

To help calibrate a turbulence model in compressible wall turbulence, a reliable transformation of compressible velocity into a universal wall law is necessary.
\citet{morkovin1962effects} hypothesizes that compressible wall-bounded flows can be mapped onto the incompressible counterparts by accounting for the variation in mean properties, as long as the turbulent Mach number is sufficiently small.
Such an argument is not entirely correct, but efforts have been made to collapse compressible velocity profiles into the incompressible universal wall law via semi-local transformation \citep{griffin2021velocity}. 
Ideally, with an accurate semi-local transformation, one can derive an explicit formula that extends a specific incompressible turbulence model into a corresponding compressible form, or just calibrate a model to match the semi-local transformation.

The history of velocity transformation dates back to the Van Driest (VD) transformation \citep{van1951turbulent} and was later improved by \cite{zhang2012mach} to enhance the collapse in the wake region of a compressible boundary layer.
However, both fail in diabatic flows.
\cite{trettel2016mean} derived an alternative to the VD transformation that works in adiabatic and diabatic channel flows (TL transformation).
A representative recent advance is the GFM transformation \citep{griffin2021velocity}, which improves the collapse of the velocity profile for compressible boundary layers in realistic fluids.
More recent proposed transformations also include \citet{hasan2023incorporating}'s transformation that suits a more general type of fluids,
and \citet{danis2024accuracy}'s approach of using eddy viscosity equivalence to collapse the velocity profiles.
There is also a data-driven approach to collapse the velocity profiles \citep{volpiani2020data}.

Many researchers have proposed compressibility corrections to turbulence models.
\citet{huang1994turbulence} argues that the classical two-equation models do not satisfy the van Driest transformation unless model constants vary as a function of the density gradient functions. 
\citet{catris2000density} developed a log-layer compressibility correction that satisfies the van Driest solution in the log layer. 
\citet{patel2018turbulence} pursued a similar approach, deriving log-layer compressibility corrections based on the semilocal scaling.
However, the corrected model by \citet{patel2018turbulence} is only applicable to channel flows, as $Re_\tau^*$ explicitly shows up in the transport equation.
The above corrections are more focused on log layers.
However, through the years of development of semi-local transformations, it has been understood that the viscous layer solution of a turbulence model also requires correction.
A very recent practice down that path is \citet{hu2025viscous}'s viscous correction to Wilcox's $k-\omega$ model \citep{wilcox1998turbulence}.
They have shown a significant improvement in channel flows, and improvements in boundary layers are also demonstrated.
However, due to the ideal of deriving an "exact" formulation based on semi-local transformation, the resulting mathematical form of their model is cumbersome.
Intuitively speaking, turbulence models are not derived exactly through the Reynolds-averaging process.
Thus, attempting to derive an exact transformation of those transport equations is unlikely to yield elegant and straightforward formulations.

An alternative approach is to explicitly correct the model by comparing it with data, as the literature contains a lot of DNS data on compressible channel flows and boundary layers.
\citet{danis2022compressibility} proposed a heuristic empirical correction, based on observing the nature of discrepancies between predictions and data.
They used local variables, such as Mach number, friction Mach number, and heat flux parameter, to form an explicit formula that multiplies a factor to the production and dissipation terms in the $\omega$ equation.
The introduction of local Mach number breaks Galilean invariance.
The model is validated in supersonic and hypersonic turbulent boundary layers with wall-cooling.

The purpose of this research is to try to find a simple, yet effective modification to a classical two-equation model, \textit{i.e.}. 
This modified $k-\omega$ model is suitable for predicting supersonic and hypersonic wall turbulence, both with and without wall cooling.
The rest of the paper is organized as follows:
\S \ref{sec:modeling} introduces the modeling philosophy for the modification, with the help of dimensional analysis and proper definition of length scales.
\S \ref{sec:channel} validated the new model in compressible turbulent channel flows.
\S \ref{sec:BL} evaluates the model in boundary layers.
\S \ref{sec:conclusion} concludes the whole work.

\section{Modeling philosophy}\label{sec:modeling}
\subsection{Revisiting the Danis-Durbin $k-\omega$ model}\label{sec:dd}
For simplicity, the Wilcox 1988 two-equation $k-\omega$ model \cite{wilcox1998turbulence} is selected as the baseline model to help clarify the rationale of the current model.
A starting point is to understand the modeling work by \citet{danis2022compressibility}, who have developed a compressibility correction for the $k-\omega$ family, including Wilcox's $k-\omega$ model and Menter's Shear Stress Transport (SST) model.
Their modifications focus on suppressing the production and dissipation terms of the $\omega$ equation, which equivalently enhances the diffusion of $\omega$ from the wall to the boundary layer, thereby reducing eddy viscosity.
The transport equation of $k$ and $\omega$ are
\begin{eqnarray}
     \frac{\partial \rho k}{\partial t}+\frac{\partial \rho k u_j}{\partial x_j}&=&\tau_{i j} \frac{\partial u_i}{\partial x_j}- C_{\mu} \rho k \omega+\frac{\partial}{\partial x_j}\left[\left(\mu+\sigma_k \frac{\rho k}{\omega}\right) \frac{\partial k}{\partial x_j}\right], \label{eqn:tke}\\ 
     \frac{\partial \rho \omega}{\partial t}+\frac{\partial \rho \omega u_j}{\partial x_j}&=& f C_{\omega 1}  \frac{\omega}{k} \tau_{i j} \frac{\partial u_i}{\partial x_j}- f C_{\omega 2} \rho \omega^2+\frac{\partial}{\partial x_j}\left[\left(\mu+\sigma_\omega \frac{\rho k}{\omega}\right) \frac{\partial \omega}{\partial x_j}\right]. \label{eqn:omega}
\end{eqnarray}
The coefficients are the same as the Wilcox's 1988 $k-\omega$ model \cite{wilcox1988reassessment}.
The only difference compared to Wilcox's 1988 formulation \cite{wilcox1998turbulence} is the damping function $f$.
The \textit{ad hoc} function is defined as
\begin{equation}
	f \left(M_\tau^*, B_q^*, M\right)=(1-\phi) \exp \left(-c_1 M_\tau^{* 2}-c_2 B_q^*\right)+\phi \exp \left(c_3 M_\tau^*+c_4 B_q^*\right),
	\label{eqn:DDf}
\end{equation}
where $\phi=\tanh \left(\frac{3}{4} M\right)$ provides an interpolation from the viscous region to the log layer, and also comes into the computation of $c_3$ and $c_4$.
The local variables are defined as $M_\tau^*\equiv u^* / a$, with $u^*\equiv\sqrt{\tau / \rho}$, 
and $B_q^* \equiv q /\left(\rho C_p T u^*\right)$. 
Their local definition of the friction Mach number $M_\tau^*$ and heat flux parameter $B_q^*$ makes the new turbulence model an entirely local formulation, independent of wall variables.
However, the involvement of local Mach number $M$, although it can be viewed as the Mach number relative to the wall, breaks Galilean invariance.

\begin{figure}
    \centering
    \includegraphics[trim={0.5cm, 0.4cm, 0.2cm, 0.2cm}, clip, width=0.95\textwidth]{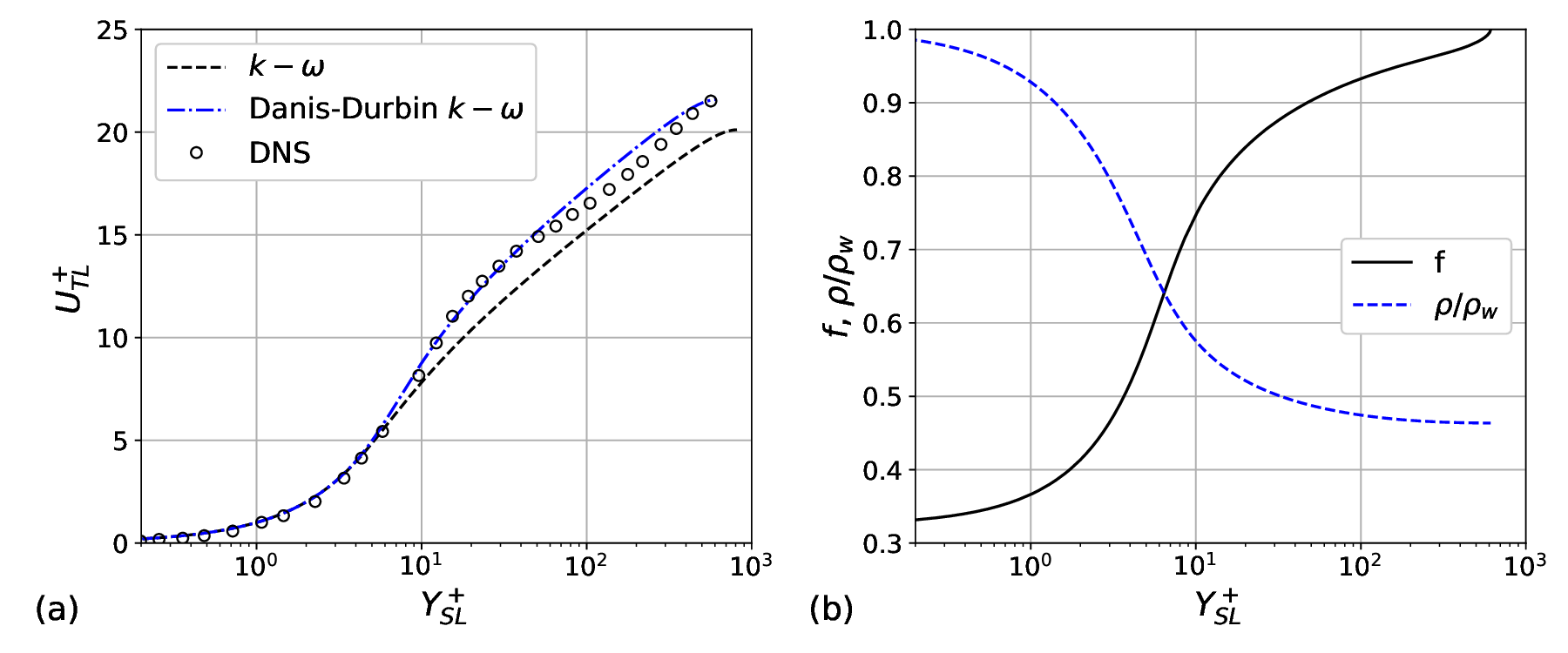}
    \caption{M3.0R600: compressible turbulent channel flow of $M_b = 3.0$, $Re_\tau^*=600$, (a) the velocity profile compared with DNS and the original $k-\omega$ model, (b) the distribution of the $f$ and normalized density $\rho/\rho_w$.}
  \label{fig:m3r600DD}
\end{figure}

Figure \ref{fig:m3r600DD} shows the performance of the Danis-Durbin $k-\omega$ model in a compressible channel flow, which is the case of M3.0R600 ($Ma_b=3.0$ and $Re_\tau^*=600$) in \citet{trettel2016mean}.
The detailed flow configuration and numerical methods will be introduced in section \S\ref {sec:channel}.
Although the model is calibrated for turbulent boundary layers, the velocity profile in Fig. \ref{fig:m3r600DD}(a) shows that the compressibility correction is effective.
The wall distance coordinate is scaled using the Trettel-Larsson transformation \cite{trettel2016mean}.
The deviation from the original $k-\omega$ model starts from the bottom of the buffer layer and extends to the center of the channel.
Figure \ref{fig:m3r600DD}(b) shows the density distribution and the distribution of $f$ inside the channel.
It shows that the $f$ value is significantly lower than unity in the viscous layer and buffer layer.
Eddy viscosity is nearly zero at $y^+ < 5$, thus the influence of $f$ on the velocity gradient in the viscous layer is nearly zero. 
Because of that, a small $f$ is not expected to play any role in the viscous layer. 
In contrast to the viscous layer, the modification to the velocity gradient in the buffer layer plays a crucial role in improving the prediction accuracy, and this is also the region where $f$ is away from unity. 
In the logarithmic layer, the velocity gradient is slightly different from the original model, and $f$ increases toward unity.
Generally speaking, the Danis-Durbin correction \cite{danis2022compressibility} is most effective in the buffer layer in terms of modifying the velocity gradient, and this is also the region with the peak density gradient.

\subsection{Representing local density effect term via dimensional analysis}
Although the Danis-Durbin $k-\omega$ model is effective, its violation of the Gealilian invariance and the involvement of many empirical functions draw attention to areas for improvement.
The analysis in \S \ref{sec:dd} also suggests that the Danis-Durbin model mostly focuses on modifying the velocity gradient where the density gradient reaches its peak, \textit{i.e.}, the buffer layer.
In light of the Morkovin hypothesis \cite{morkovin1962effects}, one might wonder whether a nondimensional gauge of local density gradient could replace the \textit{ad hoc} function of Eqn. (\ref{eqn:DDf}).
This section attempts to revise the Danis-Durbin $k-\omega$ model to achieve a similar performance, with a simpler formulation that satisfies Galilean invariance, utilizing dimensional analysis and a proper selection of length scale.

Now the problem becomes how to define this gauge of local density gradient, which is used to increase the velocity gradient properly, \textit{i.e.}, suppress Reynolds shear stress, when the wall-normal density gradient is significant.
Physically, this modification can be understood as an analogy of the phenomenon that strong, steady density stratification suppresses turbulence.
The local density gradient $\nabla \rho$ must certainly be incorporated.
Nondimensionalization requires a division by local density $\rho$.
However, $\nabla \rho/\rho$ is a vector, and only the wall-normal component needs to be considered.
Thus, $\frac{1}{\rho} \nabla \rho \cdot \nabla d$ results in a scalar indicator for the density variation normal to the wall, where $d$ is the wall distance.
The term $\frac{1}{\rho} \nabla \rho \cdot \nabla d$ has to be multiplied by a length scale $\ell$ to be dimensionless.

Since the present modeling work relates the local turbulence variable to wall cooling, the wall distance naturally comes into sight.
Additionally, the length scale should also represent the scale of local eddies that are influenced by the local density variation.
The mixing length concept may provide a choice of $\ell=\kappa d$, where $\kappa$ is the Von K\`arm\`an constant.
Since the wall distance and the mixing length only have a difference in the coefficient, and they are effectively equivalent after the final model coefficient is calibrated.
The representative length scale can be selected as the wall distance $\ell = d$.
Now, a nondimensional gauge of local wall-normal density gradient can be expressed as the following Eqn. (\ref{eqn:gauge}). 
\begin{equation}
    \alpha(\rho) = \frac{d}{\rho} (\nabla \rho \cdot \nabla d).\label{eqn:gauge}
\end{equation}

\begin{figure}
    \centering
    \includegraphics[trim={0.5cm, 0.4cm, 0.2cm, 0.2cm}, clip, width=0.95\textwidth]{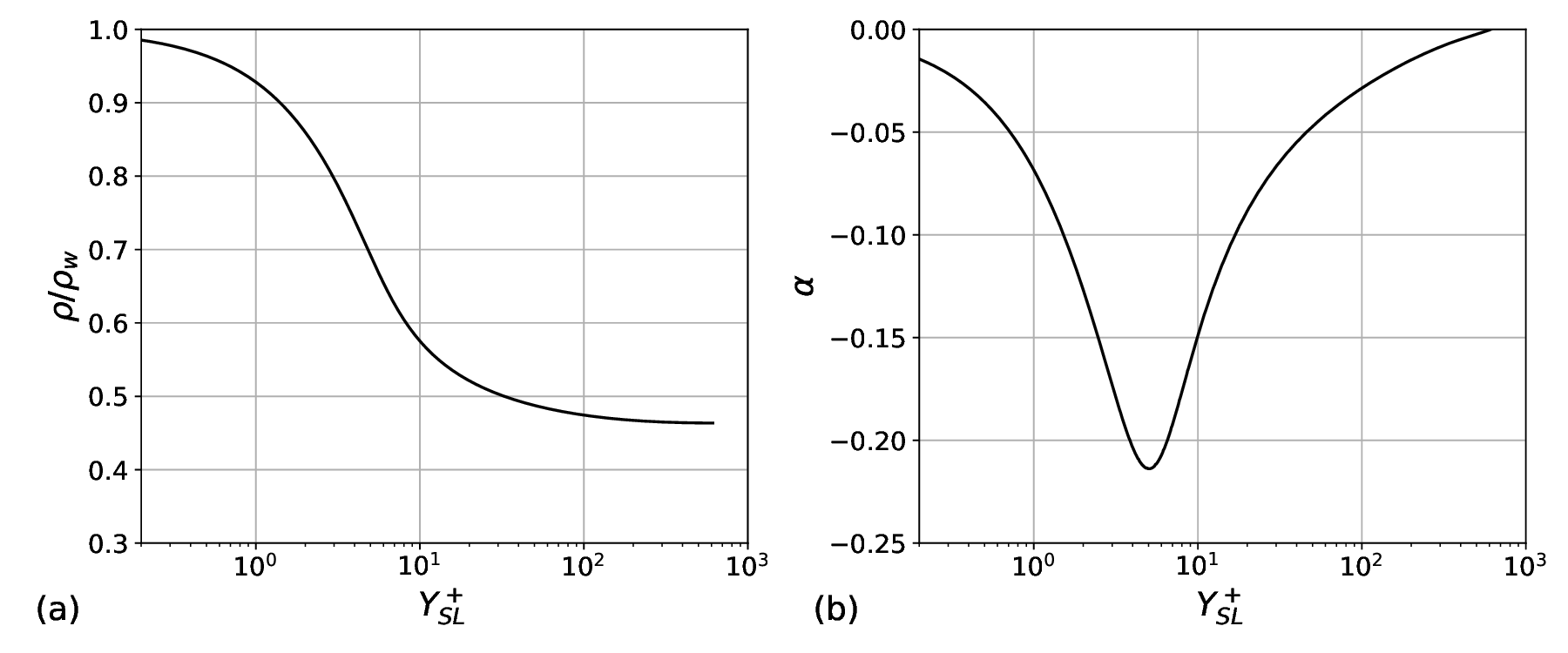}
    \caption{M3.0R600: compressible turbulent channel flow of $M_b = 3.0$, $Re_\tau^*=600$, (a) the density variation, (b) the distribution of the $\alpha$ variable.}
  \label{fig:m3r600source}
\end{figure}

To show how this nondimensional gauge for the local wall-normal density gradient works, the density field in the compressible channel flow M3.0R600 \cite{trettel2016mean}, as shown in Fig. \ref{fig:m3r600source}(a), is fed into Eqn. (\ref{eqn:gauge}).
Figure \ref{fig:m3r600source}(b) shows the resulting $\alpha(\rho)$ distribution.
Note that the distribution is different from the $f$ variable in Fig. \ref{fig:m3r600DD}(b): in the near-wall part of the viscous layer, the $\alpha$ reduces its magnitude, unlike $f$. 
What's encouraging is that the gauge function successfully detects the region where the density gradient becomes significant, and this is the region where the velocity gradient should be altered.

Now that the non-dimensional indicator of compressibility correction has been constructed into $\alpha(\rho)$, the next step is to incorporate it into a classical turbulence model, to fix the velocity profile.
Again, the two-equation Wilcox 1988 $k-\omega$ model is selected as the baseline model, as recent studies have focused on it \citep{danis2022compressibility,hu2025viscous}.
The modifications focus on suppressing the production and dissipation terms of the $\omega$ equation, which equivalently enhances the diffusion of $\omega$ from the wall to the boundary layer, thereby reducing eddy viscosity.
The approach has been verified to be effective in \citet{danis2022compressibility}, and is inherited by the current approach.

The damping function, in Eqn. \ref{eqn:omega} is given by
\begin{equation}
    f = e^{(C \times min(\alpha(\rho),0) )}. \label{eqn:f}
\end{equation}
The choice of exponential function is empirical, based on the argument that the compressible correction should vanish when $\alpha = 0$.
The empirical constant $C = 2.54$ is calibrated using the predicted velocity in the M3.0R600 compressible turbulent channel flow case.

\begin{figure}
    \centering
    \includegraphics[trim={0.25cm, 0.05cm, 0.25cm, 0.25cm}, clip, width=0.49\textwidth]{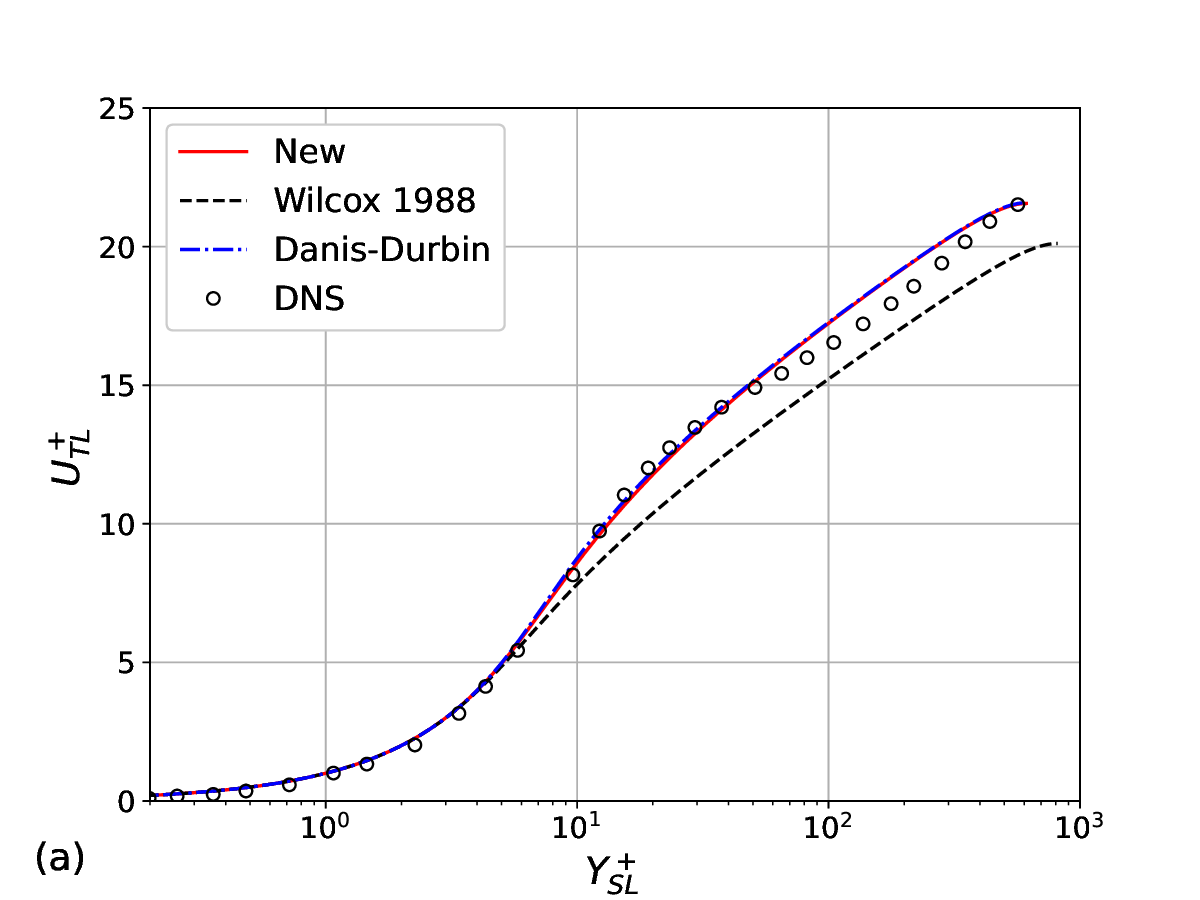}
    \includegraphics[trim={0.25cm, 0.05cm, 0.25cm, 0.25cm}, clip, width=0.49\textwidth]{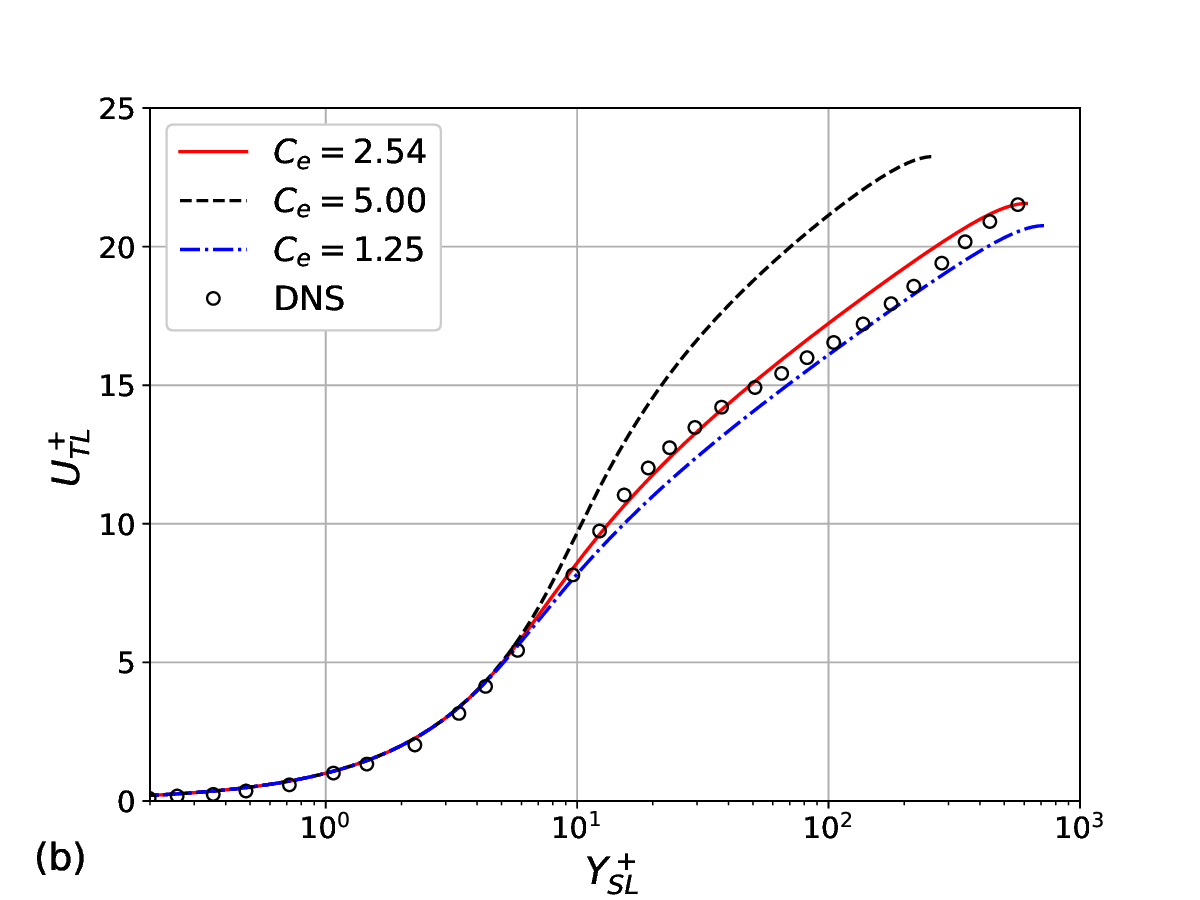}
    \caption{M3.0R600: (a) the velocity profile predicted from the new $k-\omega$ model using $C_e = 2.54$, (b) sensitivity of the velocity profile to the model constant $C_e$, compared with $C_e=1.25$ and $5.00$.}
  \label{fig:m3r600vel}
\end{figure}

The new formulation is used to simulate the channel flow M3.0R600 using $C = 2.54$.
The resulting predicted velocity profile for the M3.0R600 turbulent channel flow, after Trettel-Larsson transformation\cite{trettel2016mean}, is plotted using the red solid line in Fig. \ref{fig:m3r600vel}(a), labeled as new $k-\omega$ model.
Comparison is made with the DNS data\cite{trettel2016mean} and the simulation result using the Danis-Durbin model \cite{danis2022compressibility}, as shown in Fig. \ref{fig:m3r600vel}(a).
The new model using Eqn. (\ref{eqn:f}) produces almost identical results compared to the Danis-Durbin model \cite{danis2022compressibility} for the M3.0R600 case.
Thus, the new formulation, along with the constant $ C=2.54 $, is a good alternative to the Danis-Durbin model.
In the following sections, the Eqn. (\ref{eqn:tke}) and (\ref{eqn:omega}), with the Eqn. (\ref{eqn:f}) that enters Eqn. (\ref{eqn:omega}), will be named as the new $k-\omega$ model.

To illustrate how sensitive the new $k-\omega$ model is to the model constant $C$, simulations of the M3.0R600 turbulent channel flow are carried out using $C=1.25$ and $5.00$.
The resulting TL transformed velocity profiles are given in Fig. \ref{fig:m3r600vel}.
Therefore, the only empirical constant, $C$, controls the slope of the velocity gradient above the viscous layer.

\subsection{The complete set of control equations}
The steady-state Favre-averaged Navier–Stokes (NS) equation is solved for the following turbulent channel flows and boundary layers.
Equations (\ref{eqn:rho}) to (\ref{eqn:e}) are the solved continuity, momentum, and energy equations.
The density $\rho$ and pressure $p$ are ensemble-averaged quantities while the rest of the transported variables are Farve-averaged.
\begin{eqnarray}
\frac{\partial {\rho}}{\partial t}+\frac{\partial}{\partial x_i}\left[{\rho} {u_i}\right] & =0, \label{eqn:rho} \\
\frac{\partial}{\partial t}\left({\rho} {u_i}\right)+\frac{\partial}{\partial x_j}\left[{\rho} {u_i} {u_j}+{p} \delta_{i j}-{\tau_{i j}^{\text {tot }}}\right] & =0, \\
\frac{\partial}{\partial t}\left({\rho} {e_0}\right)+\frac{\partial}{\partial x_j}\left[{\rho} {u_j} {e_0}+{u_j} {p}+{q_j^{\text {tot }}}-{u_i} {\tau_{i j}^{\text {tot }}}\right] & =0. \label{eqn:e}
\end{eqnarray}
The stresses are defined using the Boussinesq eddy viscosity assumption,
\begin{eqnarray}
& {\tau_{i j}^{\text {tot }}} \equiv {\tau_{i j}^{\text {visc }}}+{\tau_{i j}^{\text {turb }}}, \\
& {\tau_{i j}^{\text {visc }}}=\mu\left(\frac{\partial {u_i}}{\partial x_j}+\frac{\partial {u_j}}{\partial x_i}-\frac{2}{3} \frac{\partial {u_k}}{\partial x_k} \delta_{i j}\right), \\
& {\tau_{i j}^{\text {turb }}} \approx \mu_t\left(\frac{\partial {u_i}}{\partial x_j}+\frac{\partial {u_j}}{\partial x_i}-\frac{2}{3} \frac{\partial {u_k}}{\partial x_k} \delta_{i j}\right)-\frac{2}{3} \bar{\rho} k \delta_{i j}.
\end{eqnarray}
The gradient diffusion hypothesis is used to close the heat flux terms,
\begin{eqnarray}
& {q_j^{\text {tot }}} \equiv {q_j^{\text {visc }}}+{q_j^{\text {turb }}}, \\
& {q_j^{\text {visc }}} \approx-C_p \frac{\mu}{P r} \frac{\partial {T}}{\partial x_j}, \\
& {q_j^{\text {turb }}} \approx-C_p \frac{\mu_t}{P r_t} \frac{\partial {T}}{\partial x_j}.
\end{eqnarray}

Although how to determine the turbulent Prandtl number remains an open question for compressible flows, for the current study of channel flows and boundary layers, $ \operatorname{Pr}_t = 0.9$ is used
The ideal gas law $p = \rho R T$ is used.
For the turbulent channel flow cases in \S \ref{sec:channel}, the viscosity is computed by power law following the DNS setup by \citet{trettel2016mean}.
For the turbulent boundary layers in \S \ref{sec:BL}, the viscosity is computed using the Sutherland law given in \cite{zhang2018direct,nicholson2021simulationa,nicholson2021simulationf}.

\section{Compressible turbulent channel flows}\label{sec:channel}

\subsection{Flow configurations and numerical method}
There are five compressible turbulent channel flows evaluated, and they correspond to the DNS simulation by \citet{trettel2016mean}.
A summary of simulation cases is given in Table \ref{tab:channels}.
The result of the ``M3.0R600'' case is already shown in Fig. \ref{fig:m3r600vel}.
The rest cases include two $Re_\tau^*=600$ simulations with $Ma_b = 0.7$ and $1.7$, a $Re_\tau^* = 400$ case with $Ma_b = 3.0$, and a $Re_\tau^* = 200$ case with $Ma_b = 4.0$.

\begin{table}[!htbp]
  \caption{\label{tab:channels} Simulation cases for compressible turbulent channel flow and the result $Re_\tau^*$ and temperature ratio $T_c/T_w$. DNS data from \cite{trettel2016mean}, $k-\omega$ represents Wilcox' 1988 version, Danis refers to \cite{danis2022compressibility} $k-\omega$ model, and new refers to the proposed new $k-\omega$ model.}
\centering
\def~{\hphantom{0}}
  \begin{tabular}{ l  c c c   c c c c  c c c c }
\hline
    Case  & $Ma_b$ & $Re_\tau$ & $-B_q$ & \vspace{1pt} \parbox[t]{3em}{$Re_\tau^*$\\DNS} \vspace{1pt} & \parbox[t]{3em}{$Re_\tau^*$\\Wilcox} & \parbox[t]{3em}{$Re_\tau^*$\\Danis} & \parbox[t]{3em}{$Re_\tau^*$\\New} & \parbox[t]{3em}{$T_c/T_w$\\DNS} & \parbox[t]{3em}{$T_c/T_w$\\Wilcox} & \parbox[t]{3em}{$T_c/T_w$\\Danis} & \parbox[t]{3em}{$T_c/T_w$\\New} \\
\hline
    M3.0R600 & 3.0 & 1876.1 & 0.116 & 600.7 & 814.2 & 612.9 & 611.5 & 2.49 & 1.95 & 2.45 & 2.45 \\
    M1.7R600 & 1.7 & 971.7  & 0.050 & 595.8 & 631.2 & 602.5 & 604.8 & 1.48 & 1.41 & 1.47 & 1.46 \\
    M0.7R600 & 0.7 & 652.1  & 0.010 & 591.1 & 594.8 & 594.4 & 593.6 & 1.08 & 1.08 & 1.08 & 1.08 \\
    M3.0R400 & 3.0 & 1232.5 & 0.123 & 395.5 & 543.3 & 401.2 & 402.6 & 2.49 & 1.93 & 2.46 & 2.45 \\
    M4.0R200 & 4.0 & 1017.5 & 0.189 & 202.8 & 356.5 & 237.0 & 245.4 & 3.64 & 2.32 & 3.21 & 3.12 \\
\hline
  \end{tabular}
\end{table}

Fully developed channel flows, in Reynolds-averaged form, are basically one-dimensional.
A one-dimensional solver is developed using Python to enable fast 1-D solution of the system of transport equations.
In the 1D form of the equation system, all the convection terms are zero, resulting in only source terms and diffusion terms.
The discretization of the diffusion terms is done by the 2\textsuperscript{nd} order, central finite difference scheme.
The first grid point away from the wall is located at $y^+ < 0.1$.
\revised{The $\omega$ value at the wall is set to $\omega_w=60\nu/(C_{\omega 2}y^2)$ where $y$ is the first cell center wall distance.}
Turbulent kinetic energy $k$ and eddy viscosity $\nu_t$are set to zero at the wall. 
An isothermal boundary condition is specified for temperature, and a uniform volume heat source is applied to the channel to achieve the target $B_q$ at the wall.
A pressure gradient term drives the flow to match the target $Re_\tau$.
Under-relaxation is used to converge to a steady-state solution, with a residual criterion of $\epsilon < 10^{-9}$ applied to all transport equations.

\subsection{Predicted results}
\begin{figure}
  \centering
  \begin{subfigure}{0.495\textwidth}
    \centering
    \includegraphics[trim={0.5cm, 0.25cm, 0.5cm, 0.25cm}, clip, width=0.9\textwidth]{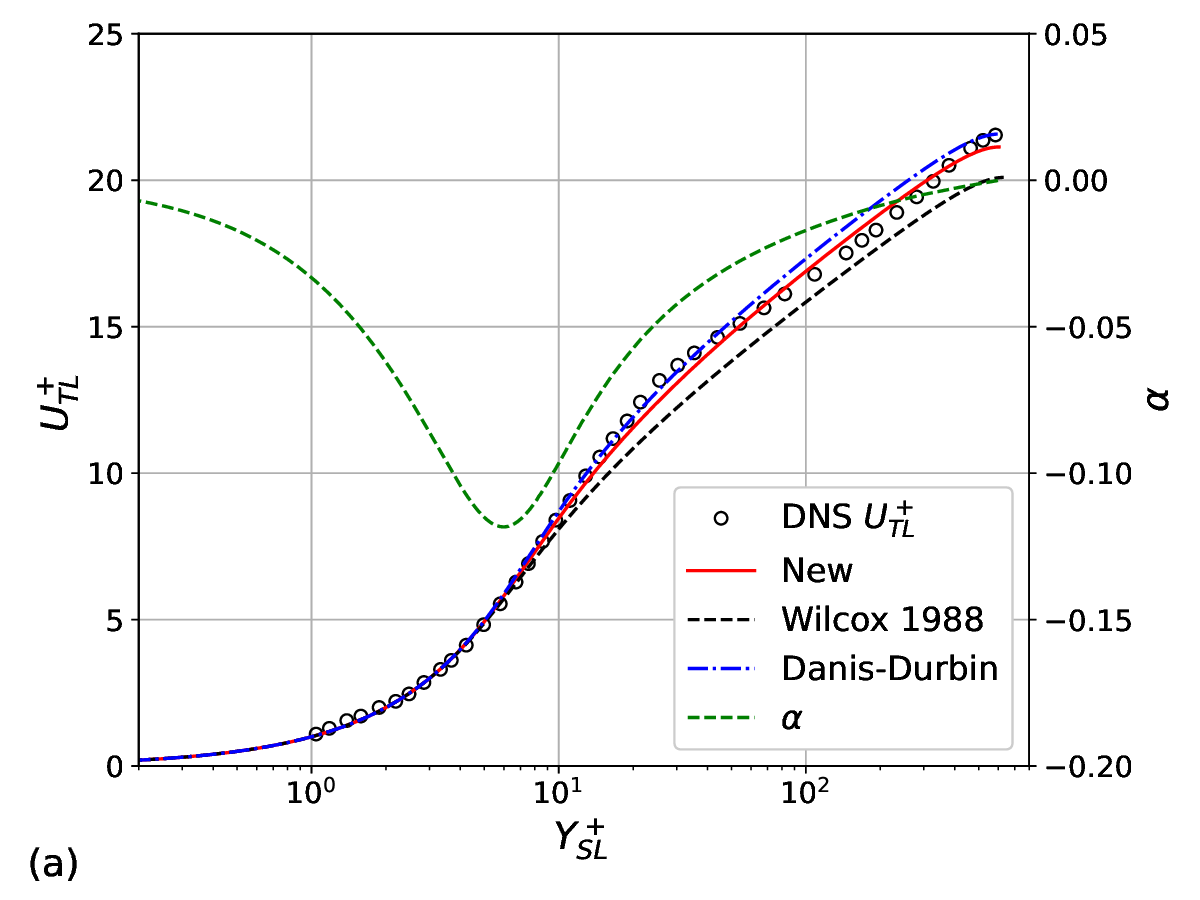}
  \end{subfigure}
  \begin{subfigure}{0.495\textwidth}
    \centering
    \includegraphics[trim={0.5cm, 0.25cm, 0.5cm, 0.25cm}, clip, width=0.9\textwidth]{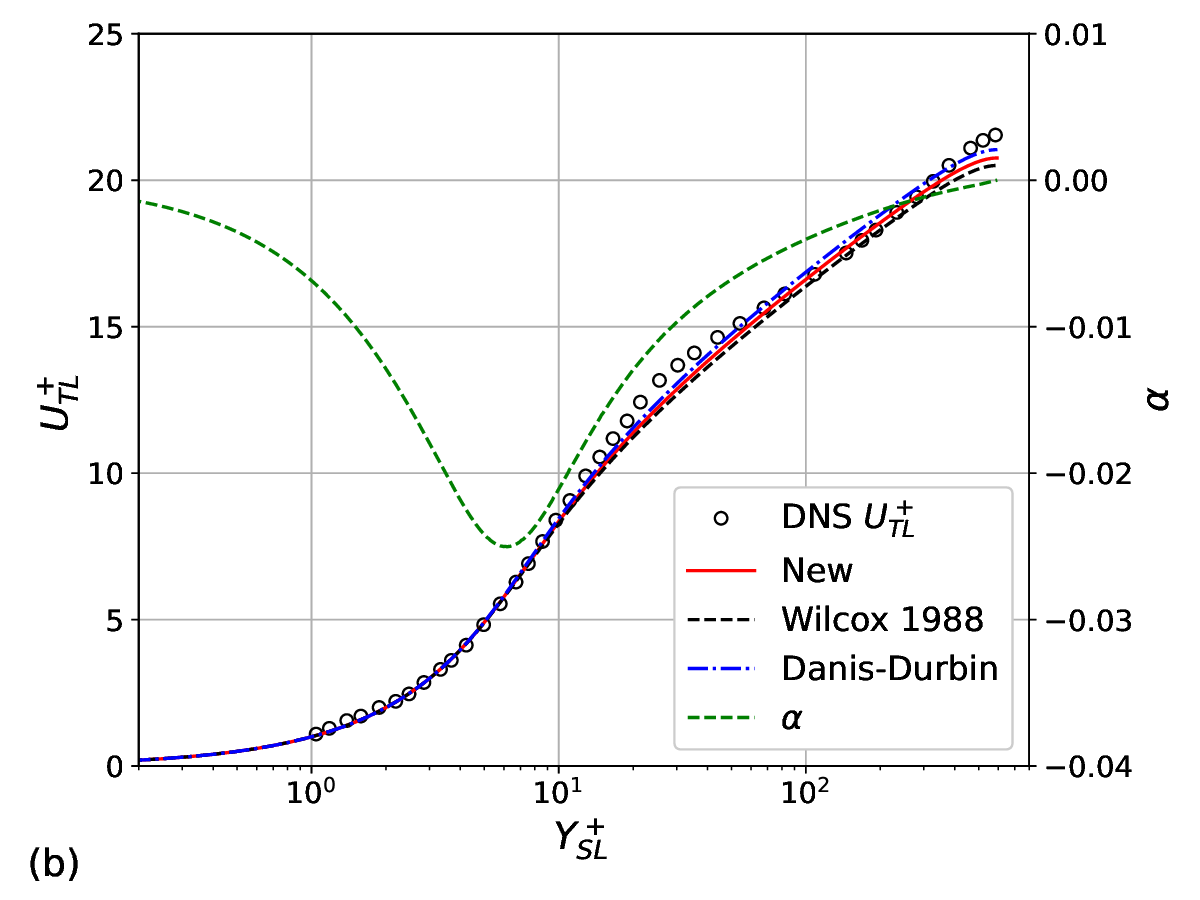}
  \end{subfigure}
    \caption{The new $k-\omega$'s predicted velocity distribution compared to the DNS data, $k-\omega$ model, and \cite{danis2022compressibility}'s model for: (a) M1.7R600, and (b) M0.7R600.}
  \label{fig:TBC1}
\end{figure}

In the previous section, the compressible channel flow M3.0R600, which corresponds to $Ma_\tau^* = 3.0$ and $Re_\tau^*=600$, is used to calibrate the model, and the predicted result is given in Fig. \ref{fig:m3r600vel}(a).
Figure \ref{fig:TBC1} shows the predicted velocity distributions at the same $Re_\tau^*=600$, but with different $Ma_\tau^*$ values of 1.7 and 0.7.
The cases are named as M1.7R600 and M0.7R600 following \citet{trettel2016mean}.
After Trettel-Larsson transformation, the predicted velocity profiles are compared with the DNS results \cite{trettel2016mean}, the original Wilcox 1988 $k-\omega$ model, and the Danis-Durbin $k-\omega$ model \citet{danis2022compressibility}.
Figure \ref{fig:TBC1} and the previous Fig. \ref{fig:m3r600vel}(a) together, demonstrate that at the same $Re_\tau^*$, the new $k-\omega$ model can automatically render a proper $\alpha$ distribution, which results in the right amount of suppression of viscosity to render the target velocity profile.
The predicted temperature ratio, computed using the channel center value $T_c$ divided by the wall value $T_w$, is also compared in Table \ref{tab:channels}.
Across M3.0R600, M1.7R600, and M0.7R600, all the $Re_\tau^*$ and the temperature ratio $T_c/T_w$ show decent agreement with the reference DNS data, suggesting that the proposed new $k-\omega$ model responded well to the Mach number effect and wall cooling effect.

\begin{figure}
  \centering
    \begin{subfigure}{0.495\textwidth}
    \centering
    \includegraphics[width=\textwidth]{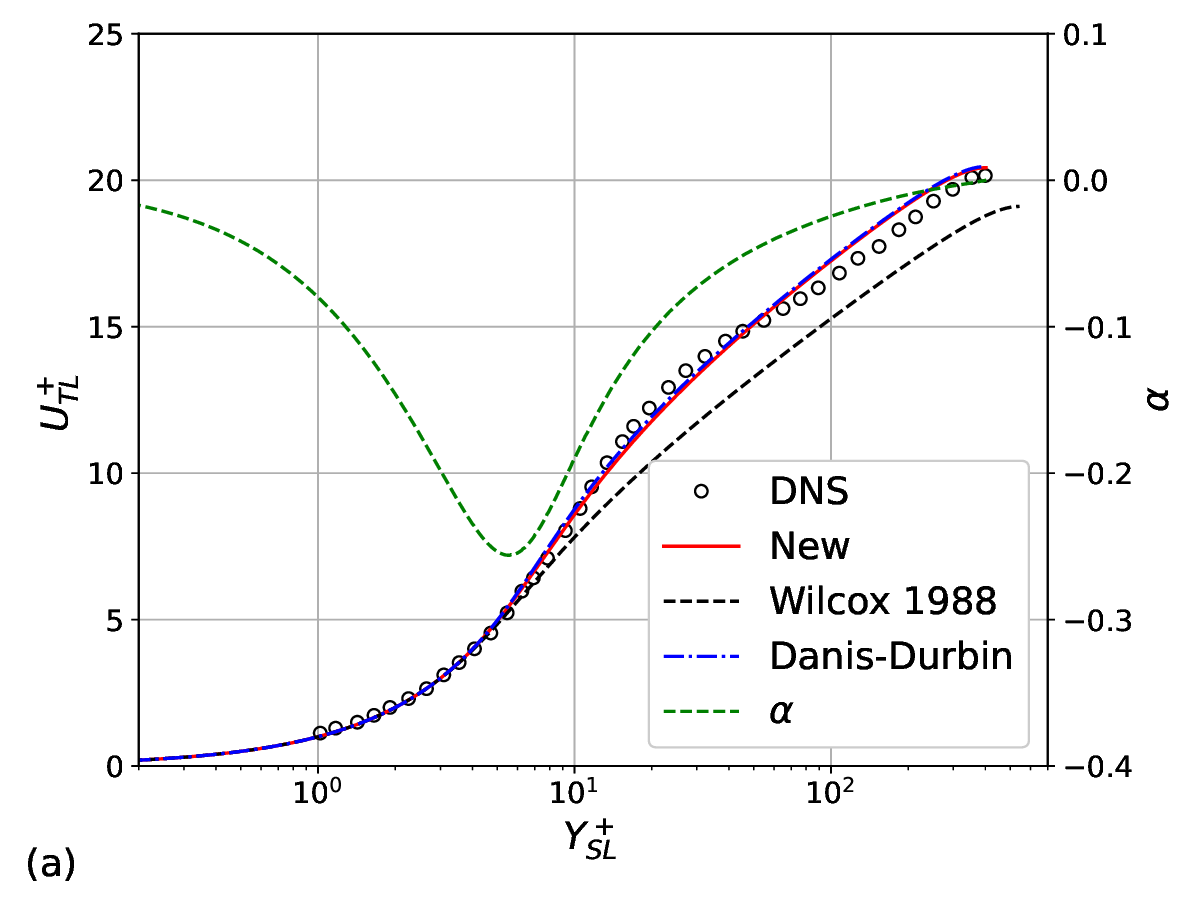}
  \end{subfigure}
  \begin{subfigure}{0.495\textwidth}
    \centering
    \includegraphics[width=\textwidth]{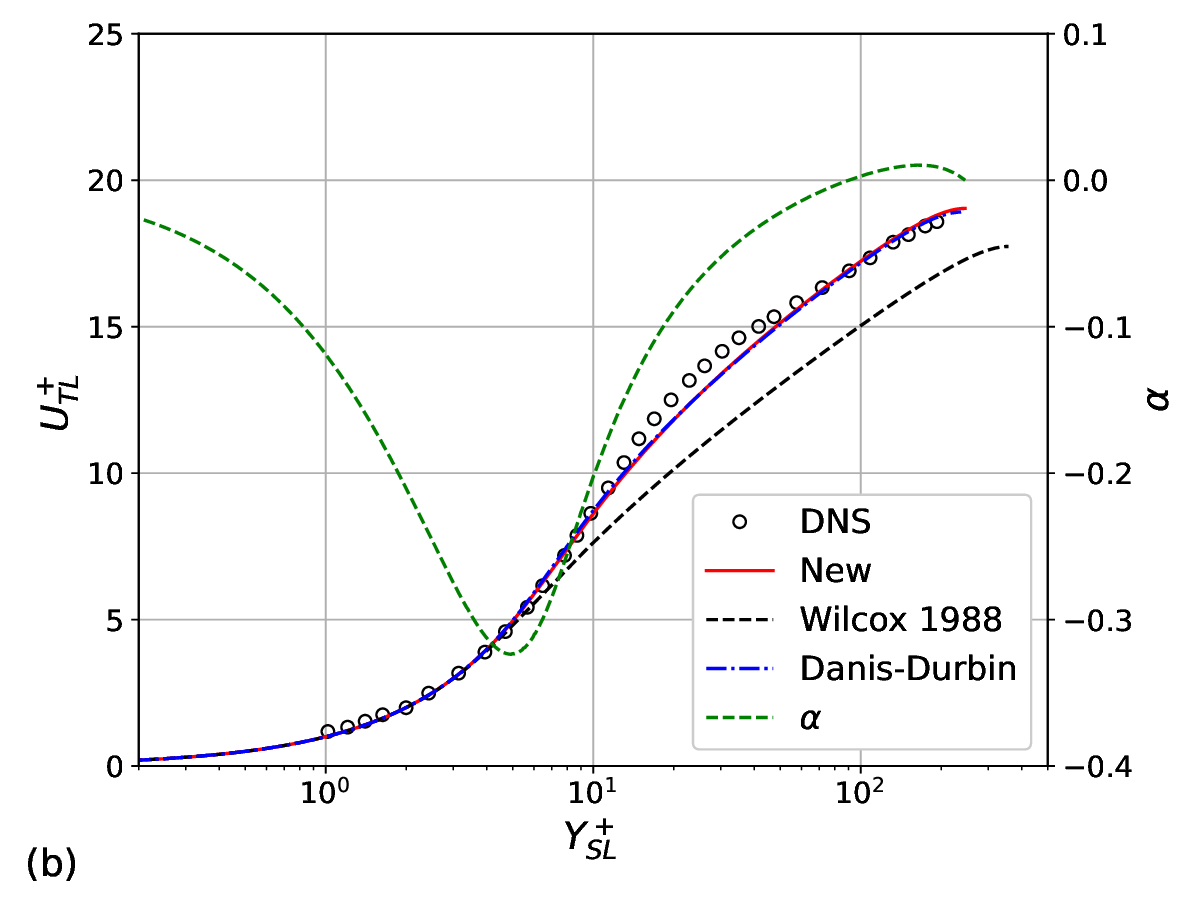}
  \end{subfigure}
    \caption{The new $k-\omega$'s predicted velocity distribution compared to the DNS data, $k-\omega$ model, and \cite{danis2022compressibility}'s model for: (a) M3.0R400, and (b) M4.0R200.}
  \label{fig:TBC2}
\end{figure}

To assess the new model's performance across different Reynolds numbers ($Re_\tau^*$), simulations are performed on the M3.0R400 and M4.0R200 cases.
Figure \ref{fig:TBC2}(a) shows the predicted velocity distributions at $Ma_b = 3.0$ and $Re_\tau^*=400$, and Figure \ref{fig:TBC2}(b) at $Ma_b = 4.0$ and $Re_\tau^*=200$.
Similar to the previous channel flow cases, the $\alpha$ distribution indicates a successful detection of the viscous and buffer layer.
The velocity profiles are significantly improved compared to the Wilcox 1988 $k-\omega$ model, and are very close to the Danis-Durbin model.
The M4.0R200 case appears to be challenging, as both $Re_\tau^*$ and $T_c/T_w$ show a larger discrepancy compared to the DNS data, when compared with other cases.
A possible reason might be the low Reynolds number effect, as the $Re_\tau^*$ is close to the lower limit of the applicable Reynolds number for a classic RANS model.

Overall, the proposed new $k-\omega$ model is capable of correcting the behavior of the original $k-\omega$ model in compressible channel flows under isothermal boundary conditions.
The model's performance is at a similar level compared to the Danis-Durbin $k-\omega$ model \cite{danis2022compressibility}.
The new model appears to be suitable for different Reynolds numbers ($Re_\tau^*$), bulk Mach numbers, and heat transfer rates.

\subsection{Scaling of turbulence variables}
A question to be answered is how the empirical modification adjusts the velocity to match the Trettel-Larsson transformation.
Semi-local transformations utilize the concept of scaling the Reynolds stresses and wall coordinate to collapse compressible profiles onto incompressible profiles.
A successful transformation would imply an eddy viscosity equivalence, $\mu_t$, in the sense that
\begin{equation}
	\frac{\mu_t}{\mu}\left(Y_{SL}^+ \right)=\frac{\mu_{t, \text { incomp }}}{\mu_w}\left(y_{+} \right).
\end{equation}

\begin{figure}
    \centering
    \includegraphics[trim={0.5cm, 0.4cm, 0.2cm, 0.2cm}, clip, width=0.99\textwidth]{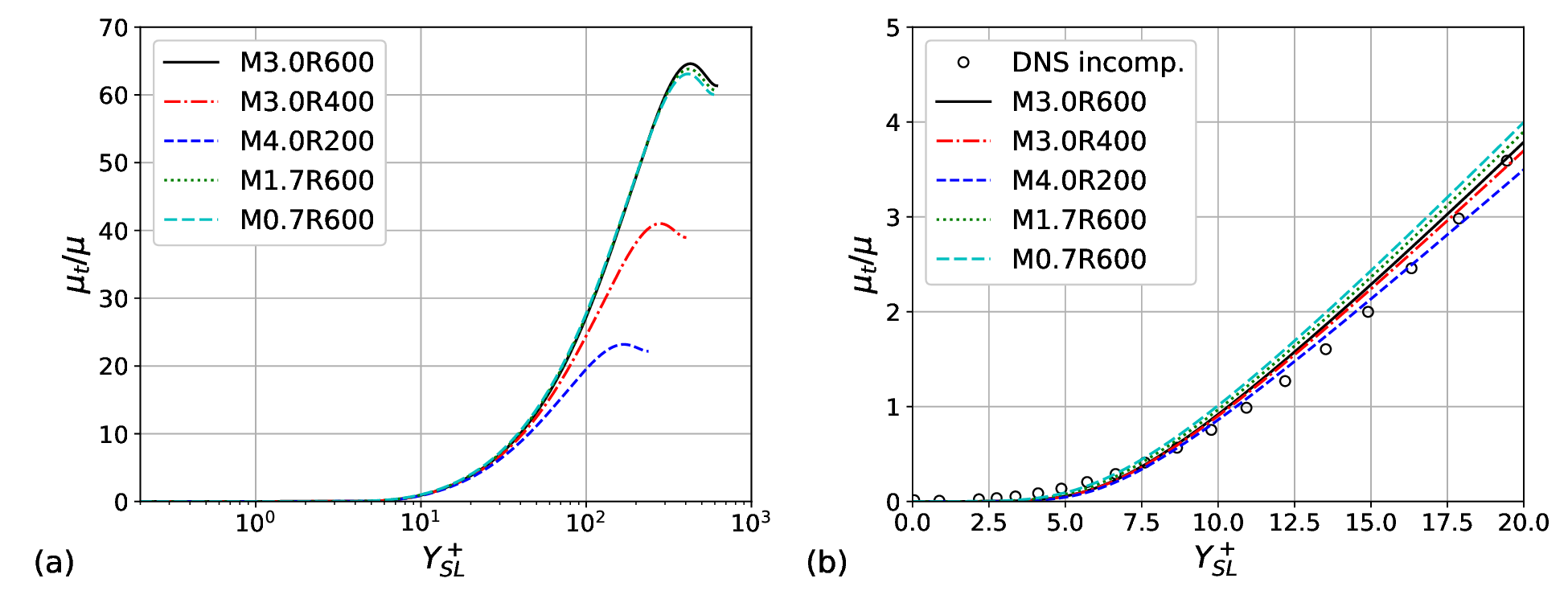}
    \caption{Scaling of turbulence variables in compressible channel flows: (a) Scaling of nondimensional eddy viscosity, (b) zoomed-in view of the nondimensional eddy viscosity.}
  \label{fig:channelScalingN}
\end{figure}

Figure \ref{fig:channelScalingN}(a) plots the turbulent eddy viscosity ratio ($\mu_t^+ \equiv \mu_t/\mu$) across all the abovementioned turbulent channel flows.
The wall distance is plotted using the Trettel-Larsson wall units.
The eddy viscosity ratio profiles collapse together in the near-wall region.
A zoomed-in view of the near-wall distribution is plotted in Fig. \ref{fig:channelScalingN}(b).
It reveals that, although slight differences remain, a satisfactory collapse onto the incompressible DNS data is observed.

\begin{figure}
    \centering
    \includegraphics[trim={0.5cm, 0.4cm, 0.2cm, 0.2cm}, clip, width=0.99\textwidth]{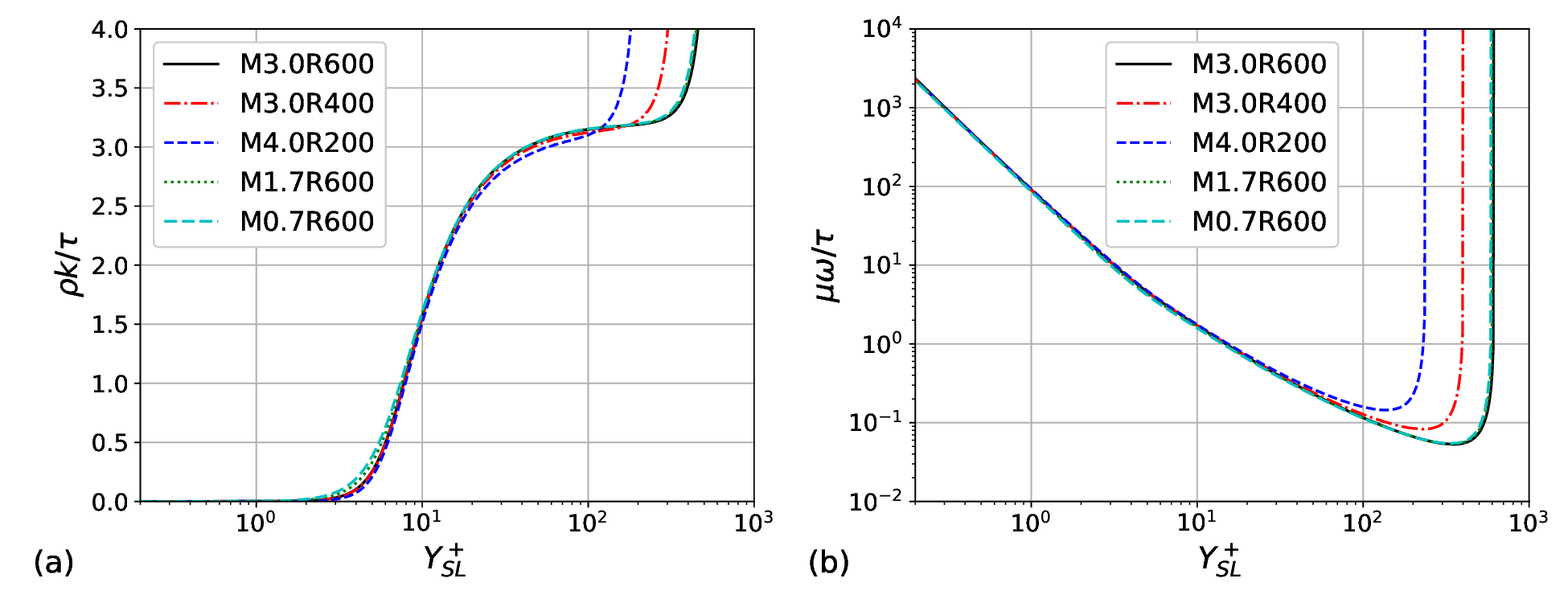}
    \caption{Scaling of turbulence variables in compressible channel flows: (a) Scaling of nondimensional turbulent kinetic energy, (b) scaling of nondimensional eddy frequency.}
  \label{fig:channelScalingTW}
\end{figure}

Given a decent improvement to the eddy viscosity distribution is observed, one may still wonder whether semi-local scalings of $k$ and $\omega$ are achieved, just by suppression of the source terms of the $\omega$ equation.
Corollaries of the previously mentioned semi-local scaling of eddy viscosity might be $\rho k=\tau G\left(Y_{SL}^+\right)$  and $\mu \omega=\tau H\left(Y_{SL}^+\right)$, where $G$ and $H$ can be found in incompressible data.
Figure \ref{fig:channelScalingTW}(a) shows the scaled $k$ and (b) the scaled $\omega$ plotted against semi-local transformed wall distance.
A very successful collapse of the eddy frequency $\omega$ is confirmed.
The collapse of turbulent kinetic energy, $k$, although not perfect, is still decent.

The examination of non-dimensional eddy viscosity, turbulent kinetic energy, and eddy frequency suggests that the proposed modification to the $k-\omega$ model is capable of correcting the original Wilcox's 1988 $k-\omega$ in isothermal compressible channel flows.
And the predicted velocity, eddy viscosity, turbulent kinetic energy, and eddy frequency profiles all collapse across different Mach number $Ma_\tau^*$ and Reynolds number $Re_\tau^*$, when the Trettel-Larsson semi-local transformation \cite{trettel2016mean} of wall coordinates is used.

\section{Compressible turbulent boundary layers with zero pressure gradient}\label{sec:BL}

Another typical wall-bounded flow is the compressible turbulent boundary layer.
Unlike channel flows, the density distribution is not monotonic within the boundary layer, which poses another challenge for the proposed model.
Zero-pressure-gradient compressible turbulent boundary layers with freestream Mach numbers ranging from 2.5 to 14.0 are selected to assess the proposed modification \cite{zhang2012mach,huang1994turbulence}.

\subsection{Numerical method and flow configurations}\label{sec:numerical}

The density-based High Speed Aerodynamic (HiSA) solver \citep{heyns2014modelling}, implemented in the open source CFD toolbox OpenFOAM-v2312 \citep{jasak2009openfoam}, is used to solve the transport equations. 
The solver uses LU-SGS with GMRES to solve continuity, momentum, and energy equations. 
Flux splitting is achieved by the AUSM+-up scheme \citep{liou2006sequel}.
Pseudo-time stepping with a local CFL number below 10 is used to converge to the steady-state solution. 
The solution process of turbulence equations is decoupled from the Navier-Stokes system.
The preconditioned Bi-conjugate gradient method is used to solve the linear systems of $k$ and $\omega$.
For the boundary conditions, a turbulent inflow intensity of 0.1\% is used to determine the inflow $k$ value.
And the inflow Dirichlet value of $\omega$ is set to make sure $\mu_{t,\infty}/\mu_{\infty} = 0.009$, as recommended by \citet{rumsey2010compressibility}.

The zero-pressure-gradient compressible turbulent boundary layers are selected from the reported DNS data \cite{zhang2012mach,huang1994turbulence}.
The freestream quantities, wall temperature, and $\delta_{99}$ at the sampling location for velocity comparison for all test cases are given in Table \ref{tab:tbl} (not including the M5 series).
The data for M11Tw020 is the DNS simulation of \citet{huang1994turbulence}, and the rest are from the DNS simulations of \citet{zhang2012mach}.
For the current 2D RANS simulation, the flow domain sizes, leading edge locations, and outflow plane positions, for all cases, are identical to those in the practice of \citet{danis2022compressibility}.
The total grid numbers in both streamwise ($N_x$) and wall-normal ($N_y$) directions are also consistent with their practice \cite{danis2022compressibility}.
Although \citet{danis2022compressibility} employs the VULCAN-CFD code that might be more appropriate than OpenFOAM for hypersonic flows, RANS is not very sensitive to the numerical dissipation of a CFD code; grid convergence is also achieved in the current study.
The grid points are stretched near the solid boundary so that the $y^+$ for the point next to the wall is below 0.6.

\begin{table}[hbt!]
\caption{\label{tab:tbl} Freestream quantities, wall-temperature, the criteria for determining the sampling location for various test cases, the computational mesh dimension, and computational domain size.}
\centering
\def~{\hphantom{0}}
  \begin{tabular}{ l c c c c c c l c c c c}
\hline
    Case     &$Ma_\infty$& $U_\infty$, m/s & $\rho_\infty$, kg/m$^3$ & $T_\infty$, K & $T_w$, K & $T_w/T_r$ & $x_{loc}$ & $N_x$ & $N_y$ & $L_x$ & $L_y$ \\
\hline
    M2p5     & 2.5       & 823.6  & 0.100 & 270.0 & 568.0 & 1.00 & $\delta$ = 7.7 mm & 350 & 200 & 0.70 & 0.164 \\
    M6Tw025  & 5.84      & 869.1  & 0.044 &  55.2 &  97.5 & 0.25 & $\delta$ = 3.6 mm & 310 & 200 & 0.31 & 0.075 \\
    M6Tw076  & 5.86      & 870.4  & 0.043 &  55.0 & 300.0 & 0.76 &$\delta$ = 23.8 mm & 345 & 200 & 2.32 & 0.550 \\
    M8Tw048  & 7.87      & 1155.1 & 0.026 &  51.8 & 298.0 & 0.48 &$\delta$ = 35.2 mm & 310 & 200 & 3.10 & 0.824 \\
    M11Tw020 &10.90      & 1778.4 & 0.103 &  66.5 & 300.0 & 0.20 & $\delta$ = 8.0 mm & 900 & 250 & 1.50 & 0.200 \\
    M14Tw018 &13.64      & 1882.2 & 0.017 &  47.4 & 300.0 & 0.18 &$\delta$ = 66.1 mm & 500 & 250 & 6.20 & 1.045 \\
\hline
    M5pWPG   & 4.86       & 794.0  & 0.272 & 66.2 & 317.0 & 0.91 & $x$ = 0.298 m & 350 & 200 & 0.9 & 0.24 \\
    M5pSPG   & 4.86       & 794.0  & 0.272 & 66.2 & 317.0 & 0.91 & $x$ = 0.298 m & 350 & 200 & 0.9 & 0.24 \\
    M5pAPG   & 4.86       & 794.0  & 0.272 & 66.2 & 317.0 & 0.91 & $x$ = 0.308 m & 350 & 200 & 0.9 & 0.24 \\
\hline
\end{tabular}
\end{table}

\subsection{Predicted results}
\begin{figure}
  \centering
  \begin{subfigure}{0.495\textwidth}
    \centering
    \includegraphics[trim={0.25cm, 0.25cm, 0.5cm, 0.35cm}, clip, width=0.9\textwidth]{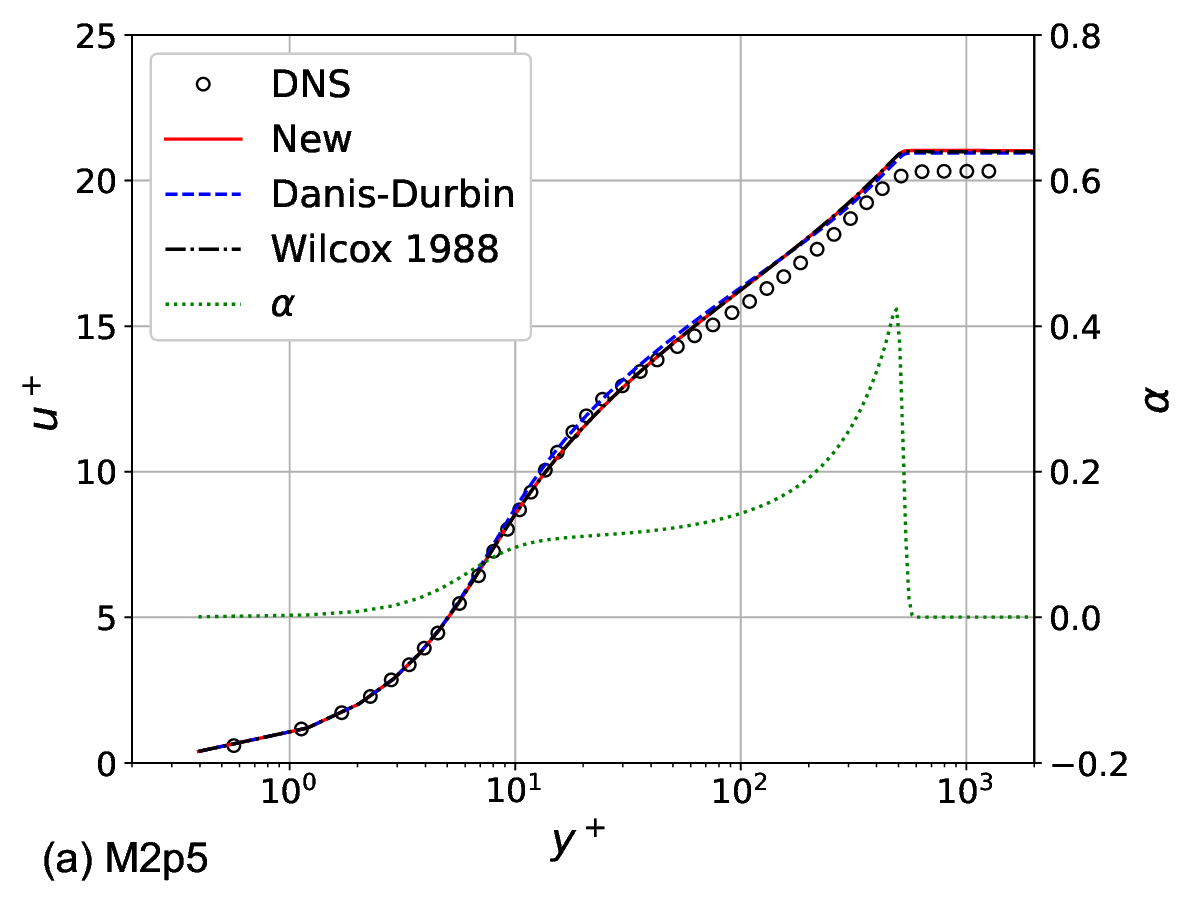}
  \end{subfigure}
  \begin{subfigure}{0.495\textwidth}
    \centering
    \includegraphics[trim={0.25cm, 0.25cm, 0.5cm, 0.35cm}, clip, width=0.9\textwidth]{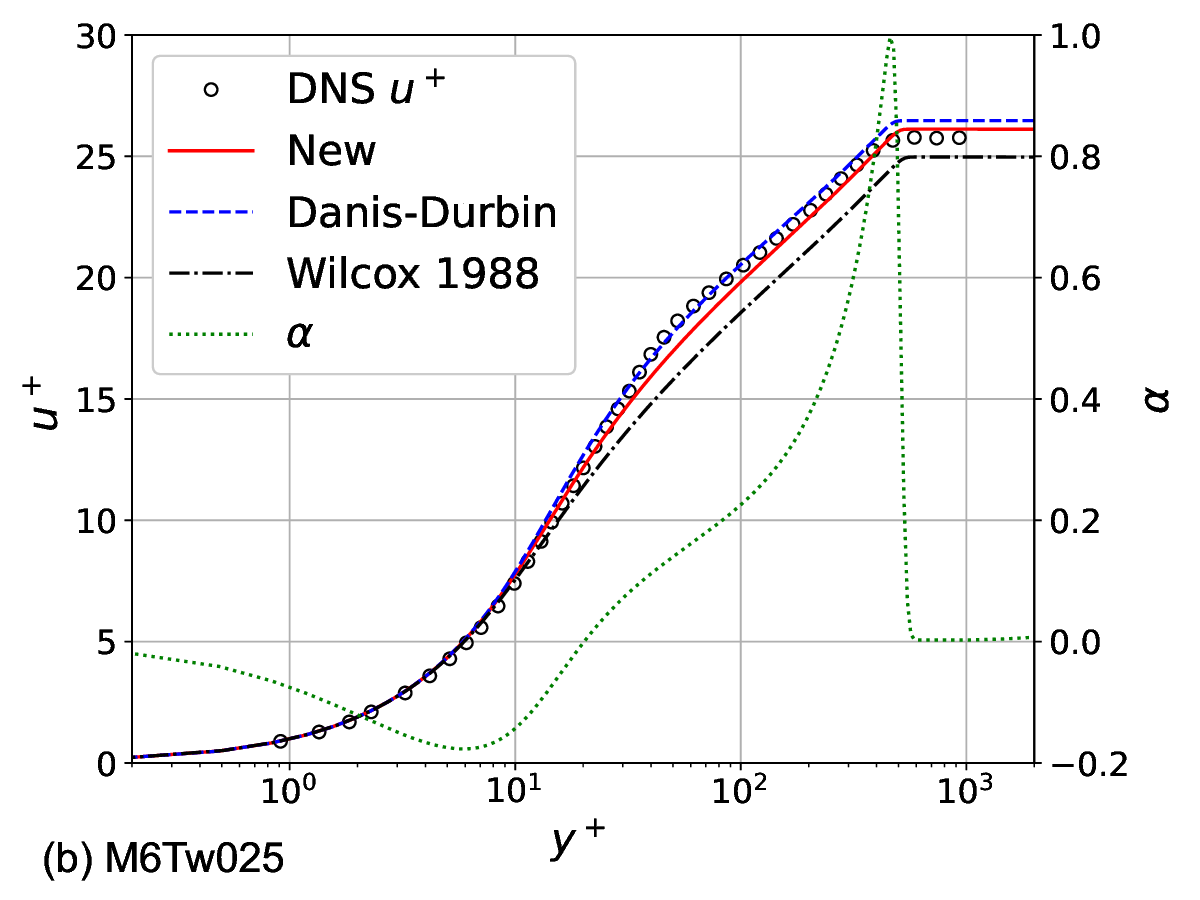}
  \end{subfigure}
  \begin{subfigure}{0.495\textwidth}
    \centering
    \includegraphics[trim={0.25cm, 0.25cm, 0.5cm, 0.35cm}, clip, width=0.9\textwidth]{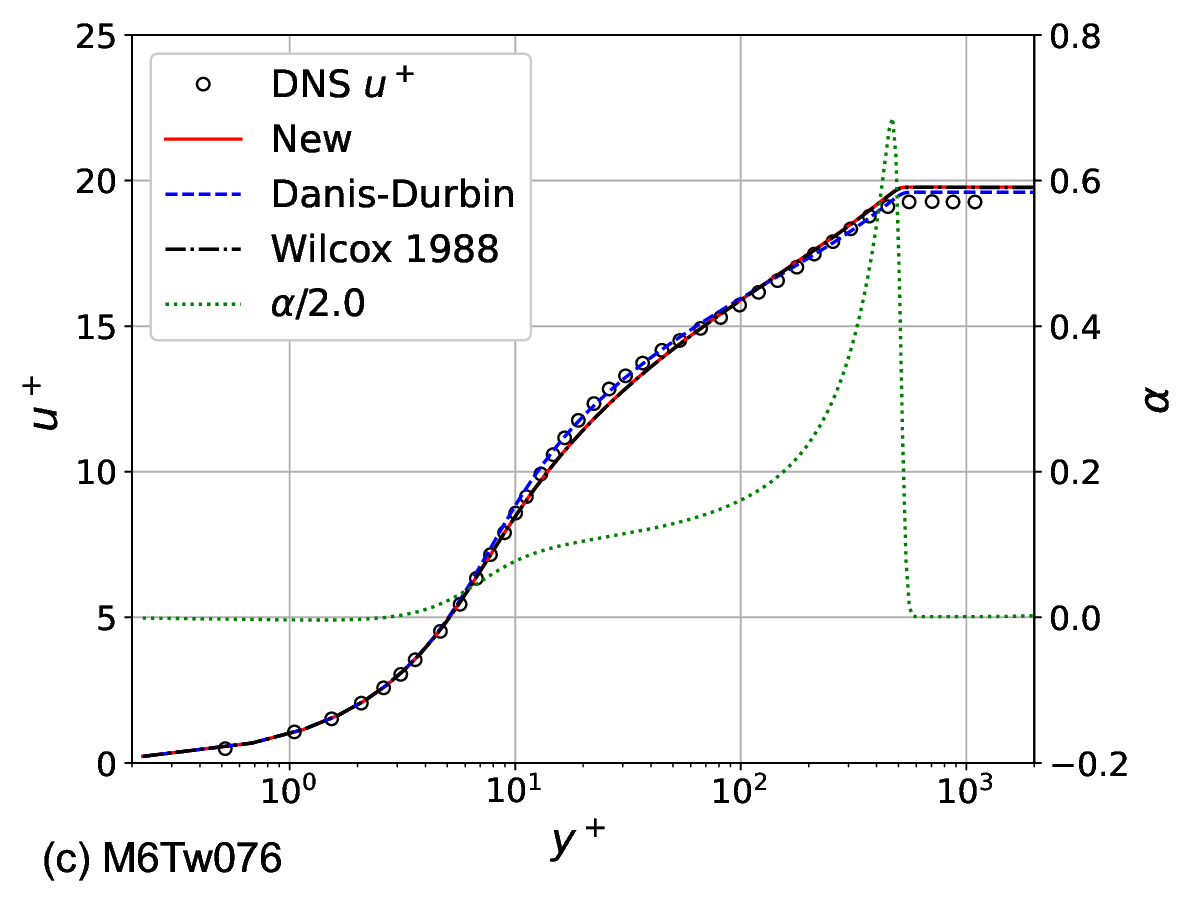}
  \end{subfigure}
  \begin{subfigure}{0.495\textwidth}
    \centering
    \includegraphics[trim={0.25cm, 0.25cm, 0.5cm, 0.35cm}, clip, width=0.9\textwidth]{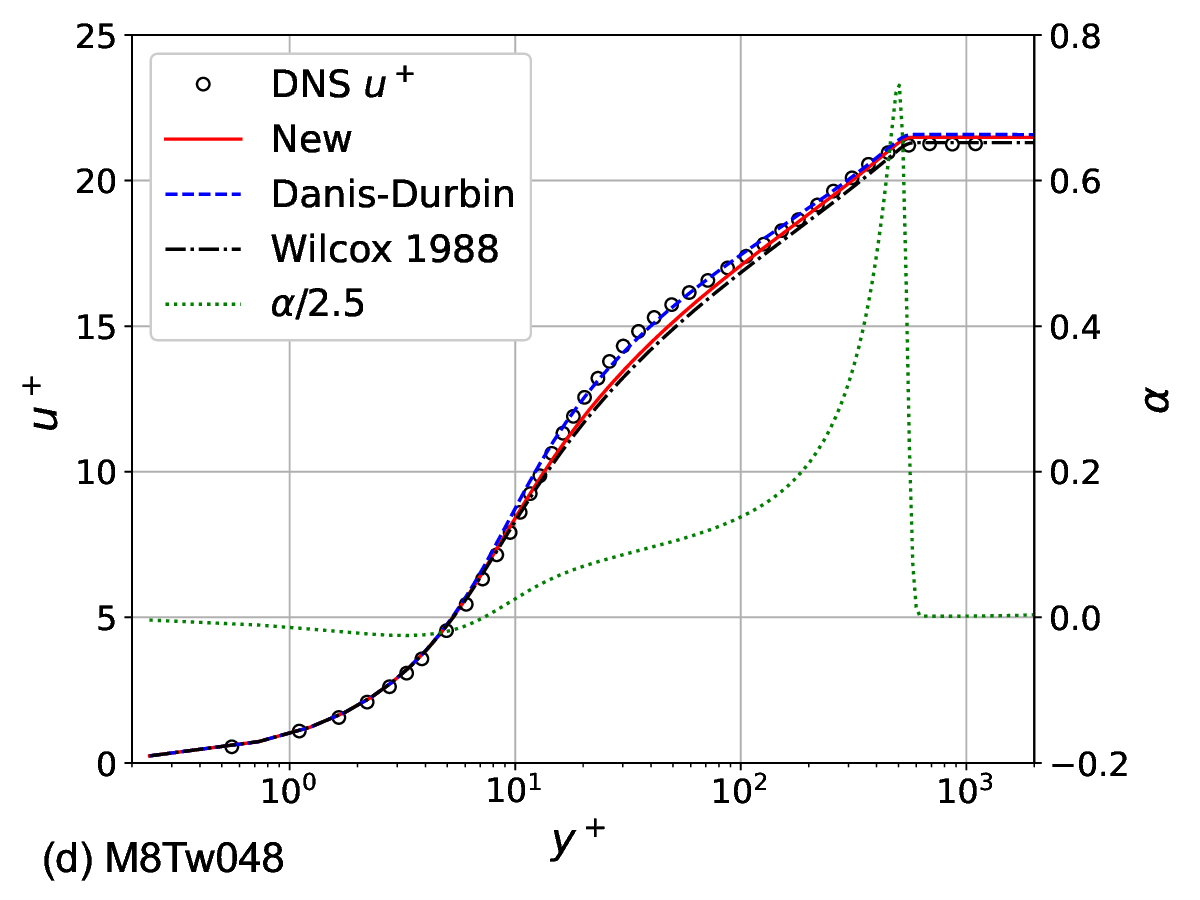}
  \end{subfigure}
  \begin{subfigure}{0.495\textwidth}
    \centering
    \includegraphics[trim={0.25cm, 0.25cm, 0.5cm, 0.35cm}, clip, width=0.9\textwidth]{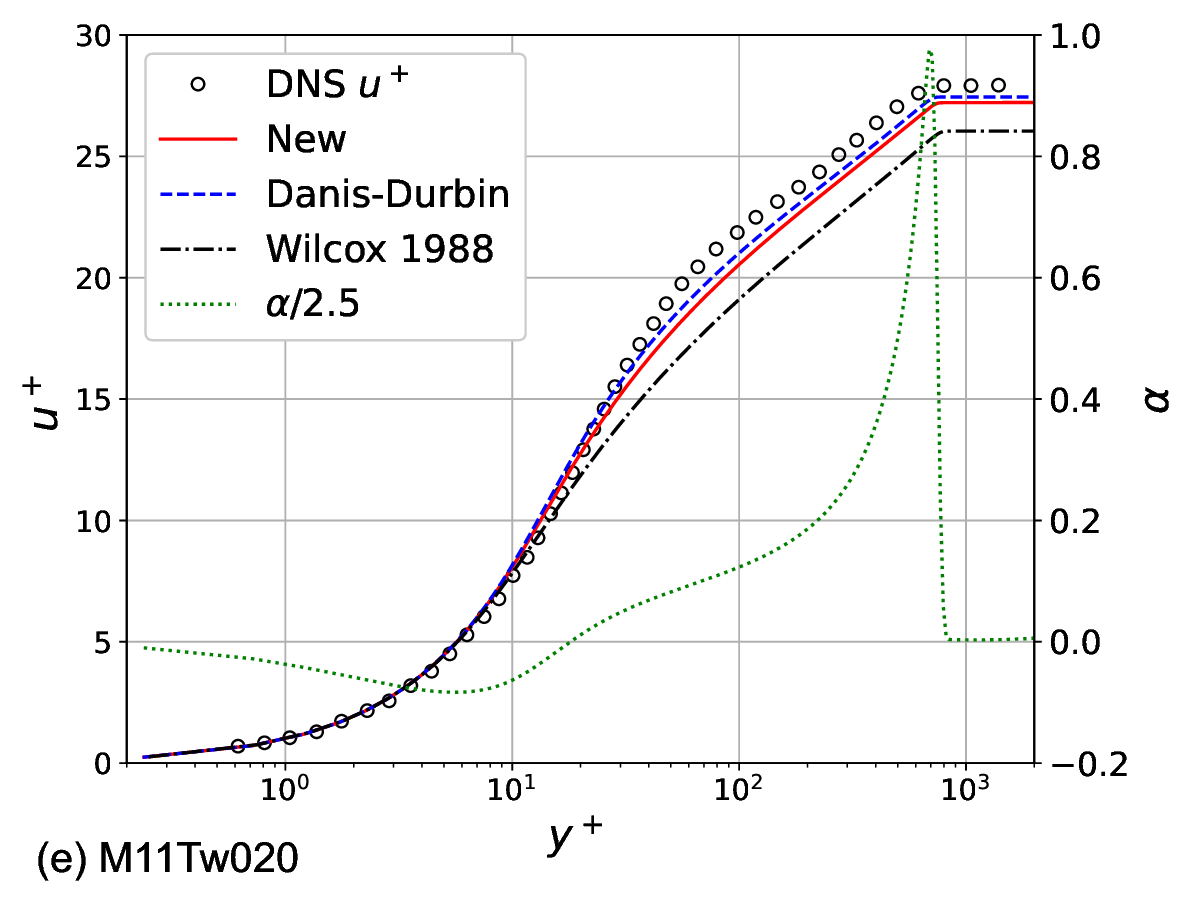}
  \end{subfigure}
  \begin{subfigure}{0.495\textwidth}
    \centering
    \includegraphics[trim={0.25cm, 0.25cm, 0.5cm, 0.35cm}, clip, width=0.9\textwidth]{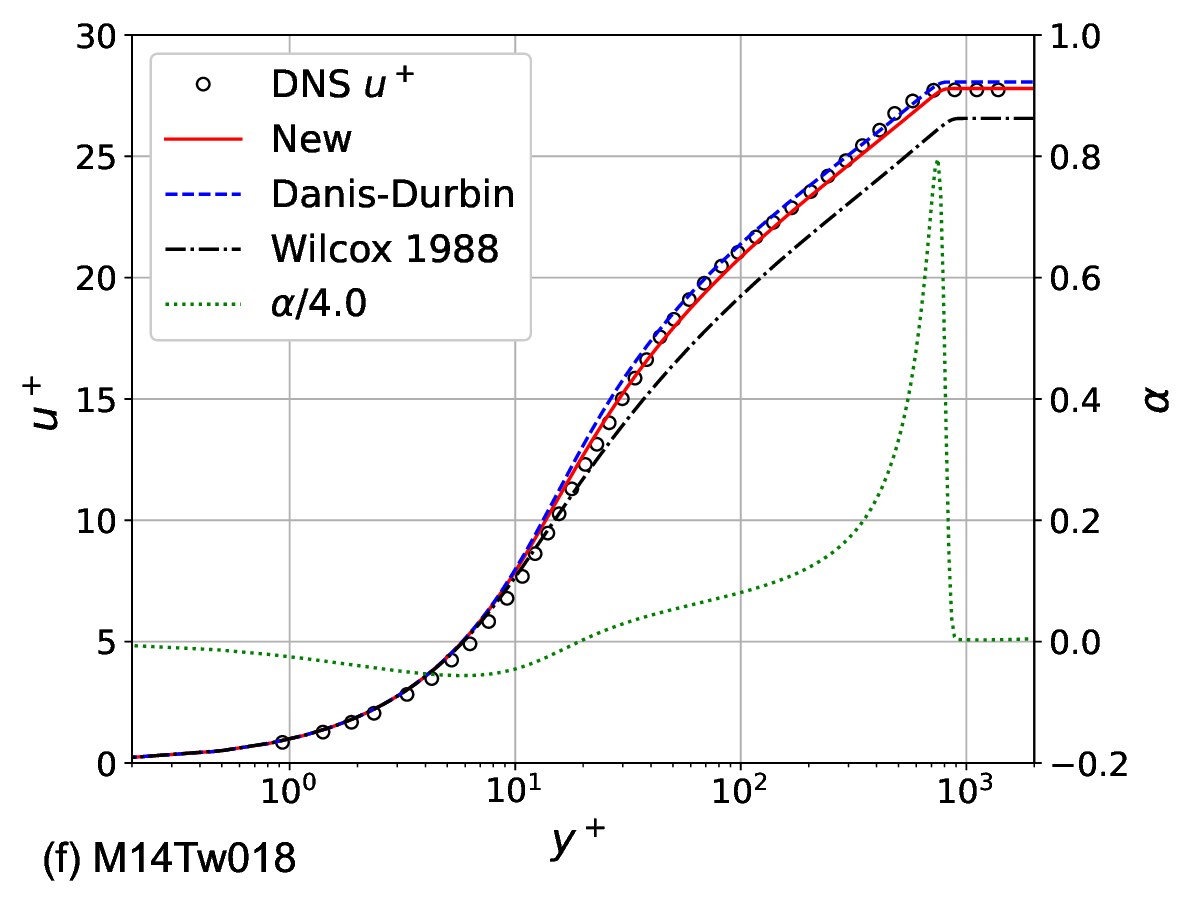}
  \end{subfigure}
  \caption{The new $k-\omega$'s predicted nondimensional velocity distribution compared to the DNS data and Wilcox's $k-\omega$ model, and the distribution of $\alpha$. The cases are (a) M2p5, (b) M6Tw025, (c) M6Tw076, (d) M8Tw048, (e) M11Tw020, and (f) M14Tw018.}
  \label{fig:TBLV}
\end{figure}

The new $k-\omega$ model's predicted velocity distributions, normalized by wall quantities, together with the distribution of $\alpha$, are plotted in Fig. \ref{fig:TBLV}.
Following \citet{danis2022compressibility}'s practice, the velocity and wall distance are normalized by wall quantities directly.
No semi-local transformation is applied in the plotting, so that the prediction error is not interfered with by the specified choice of transformation.
Clearly, the distribution of $\alpha$ is quite different compared to channel flow cases.
In the reasoning of applying $\alpha$ to correct the $k-\omega$ model, only suppression of eddy viscosity in the near wall region is desired.
Thus, the positive value of $\alpha$, near the top of the boundary layer, is not expected to interfere with the $k-\omega$ equations.
This explains why an upper limit of 0 for $\alpha$ is introduced into Eqn. (\ref{eqn:f}).
Considering the differences between the Wilcox 1988 $k-\omega$, the Danis-Durbin $k-\omega$ model, the new $k-\omega$, and the DNS in predicting the evolution of the boundary layer, the comparisons are made at the sampling locations with the same $\delta_{99}$.

For the cases of (a) M2p5 and (c) M6Tw076, the performance of the three models is expected to be similar, as there is no wall-cooling in (a) M2p5 and very weak in (c) M6Tw076. 
For the two cases, the profiles of $\alpha$ show all positive values throughout the entire boundary layer, and become zero in the freestream.
A consequence of the positive $\alpha$ profile is the resulting uniform distribution of $f=1$.
Thus, for the (a) M2p5 and (c) M6Tw076 cases, the new $k-\omega$ model is identical to the Wilcox 1988 $k-\omega$ model. 
The Danis-Durbin $k-\omega$ model \cite{danis2022compressibility} yields results that are very similar to those of the Wilcox 1988 $k-\omega$ model and the new model.
The performance of the three models is expected to be similar, as there is no wall-cooling in (a) M2p5 and very weak in (c) M6Tw076.

When wall cooling is mild, \textit{e.g.}, (d) M8Tw048, a negative value of $\alpha$ is observed below $y^+=10$.
However, the predicted velocity from the new model is very similar to Wilcox's $k-\omega$ model.
The Danis-Durbin model, on the other hand, shows improvement in the buffer layer and the bottom of the log layer.
In terms of the freestream $u^+$, which reflects the accuracy on the $u_\tau$ and skin friction, the new model and Wilcox's $k-\omega$ model are slightly better than the Danis-Durbin model.

Under the condition of strong wall cooling, which is the case for (b) M6Tw025, (e) M11Tw020, and (f) M14Tw018, significant improvement to Wilcox's $k-\omega$ model is observed for both the Danis-Durbin model and the new model.
The Danis-Durbin model is more aggressive than the new model.
The $\alpha$ distribution exhibits an obvious negative portion that begins in the viscous layer and extends into the buffer layer.
The improvement in the velocity profile is primarily attributed to the change in the velocity gradient in the buffer layer. 
The normalized $u^+$ values in the freestream for M6Tw025 and M14Tw018 are much closer to the DNS data compared to Wilcox's $k-\omega$ model, and are slightly better than the Danis-Durbin model.
An accurate prediction of $u^+$ in the freestream indicates a very decent prediction of the wall shear stress.
However, for M11Tw020, the new $k-\omega$ model's improvement, although still encouraging, is slightly weaker than that of the Danis-Durbin model.

Generally speaking, the prediction results in Fig. \ref{fig:TBLV} suggest that the new $k-\omega$ model can certainly improve boundary layer velocity prediction under mild and strong wall-cooling effects.
And, if the wall cooling is weak or nonexistent, the new model reverts to the original Wilcox's $k-\omega$ model.

\begin{figure}
  \centering
  \begin{subfigure}{0.495\textwidth}
    \centering
    \includegraphics[trim={0.5cm, 0.25cm, 0.35cm, 0.25cm}, clip, width=0.9\textwidth]{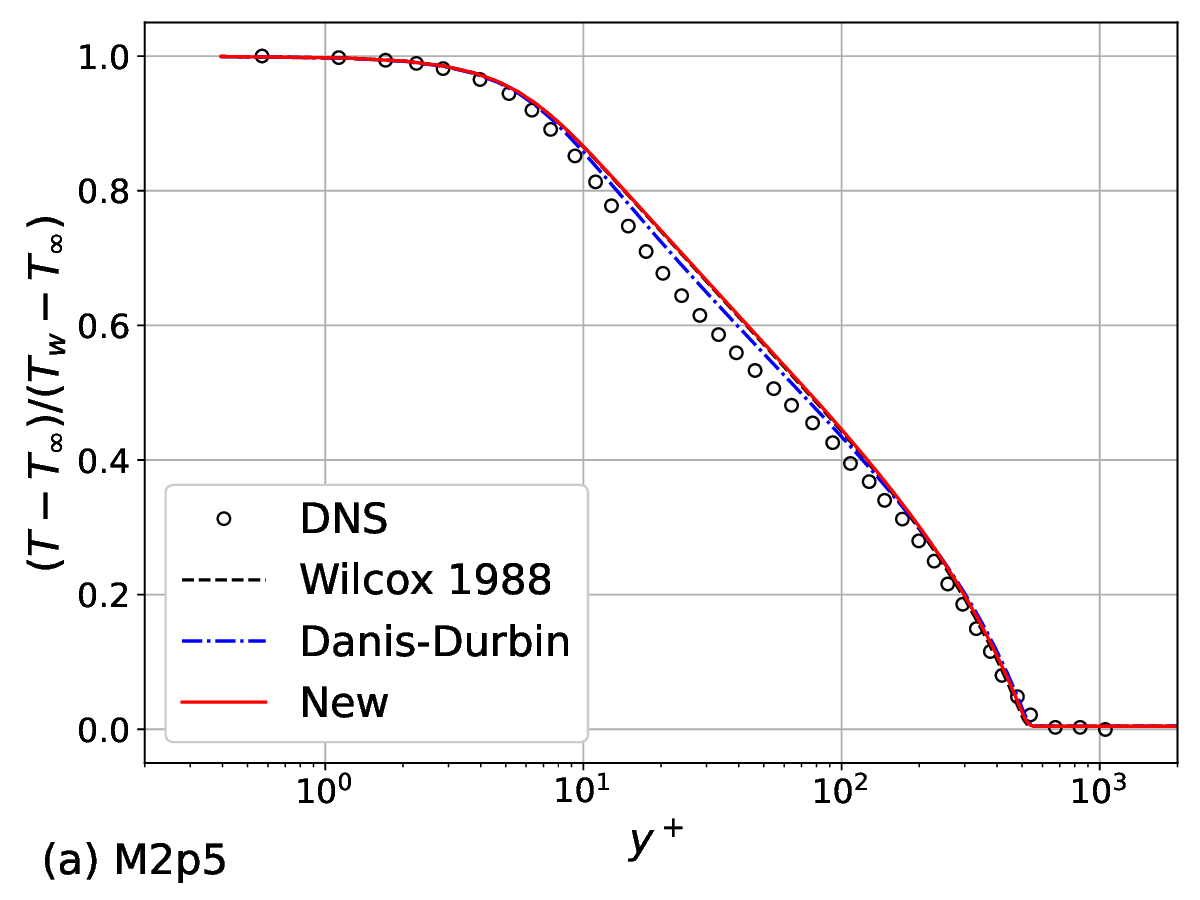}
  \end{subfigure}
  \begin{subfigure}{0.495\textwidth}
    \centering
    \includegraphics[trim={0.5cm, 0.25cm, 0.35cm, 0.25cm}, clip, width=0.9\textwidth]{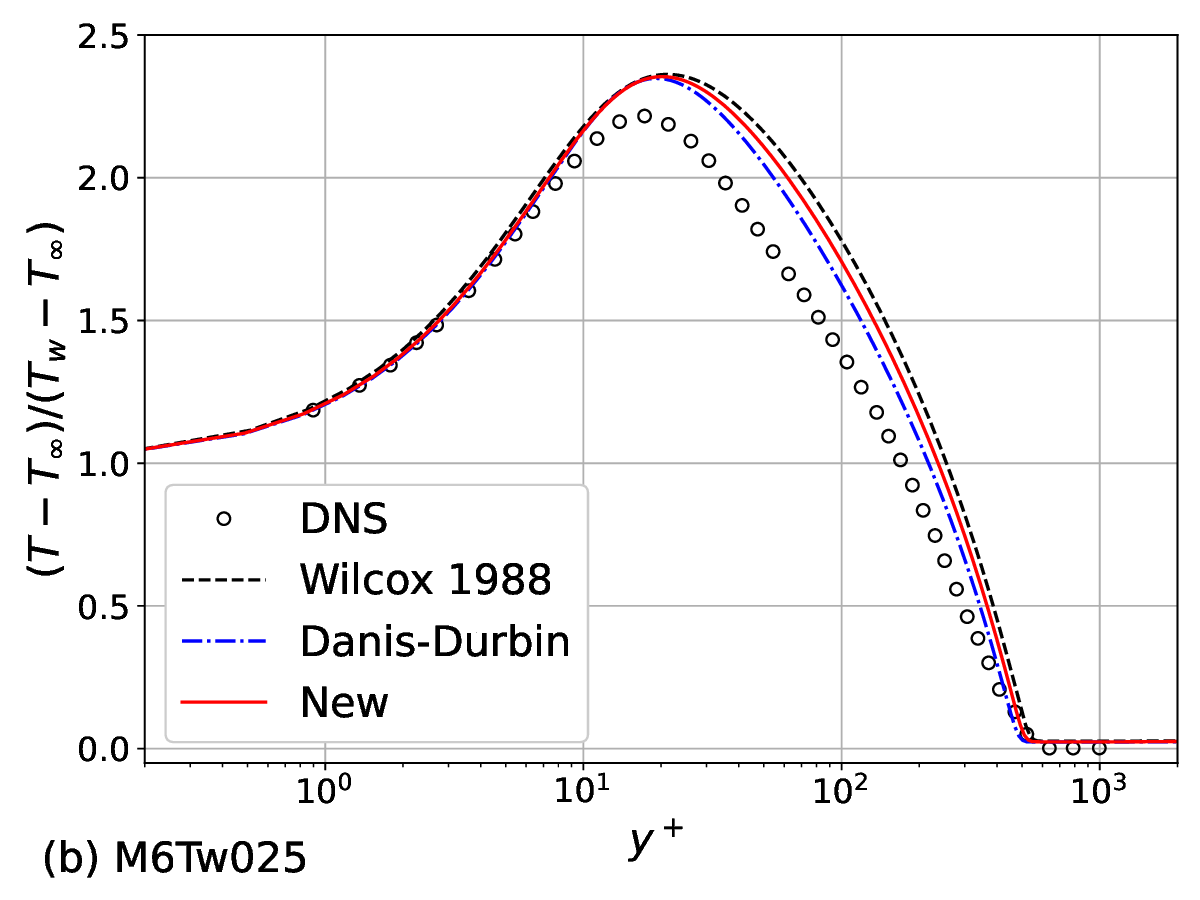}
  \end{subfigure}
  \begin{subfigure}{0.495\textwidth}
    \centering
    \includegraphics[trim={0.5cm, 0.25cm, 0.35cm, 0.25cm}, clip, width=0.9\textwidth]{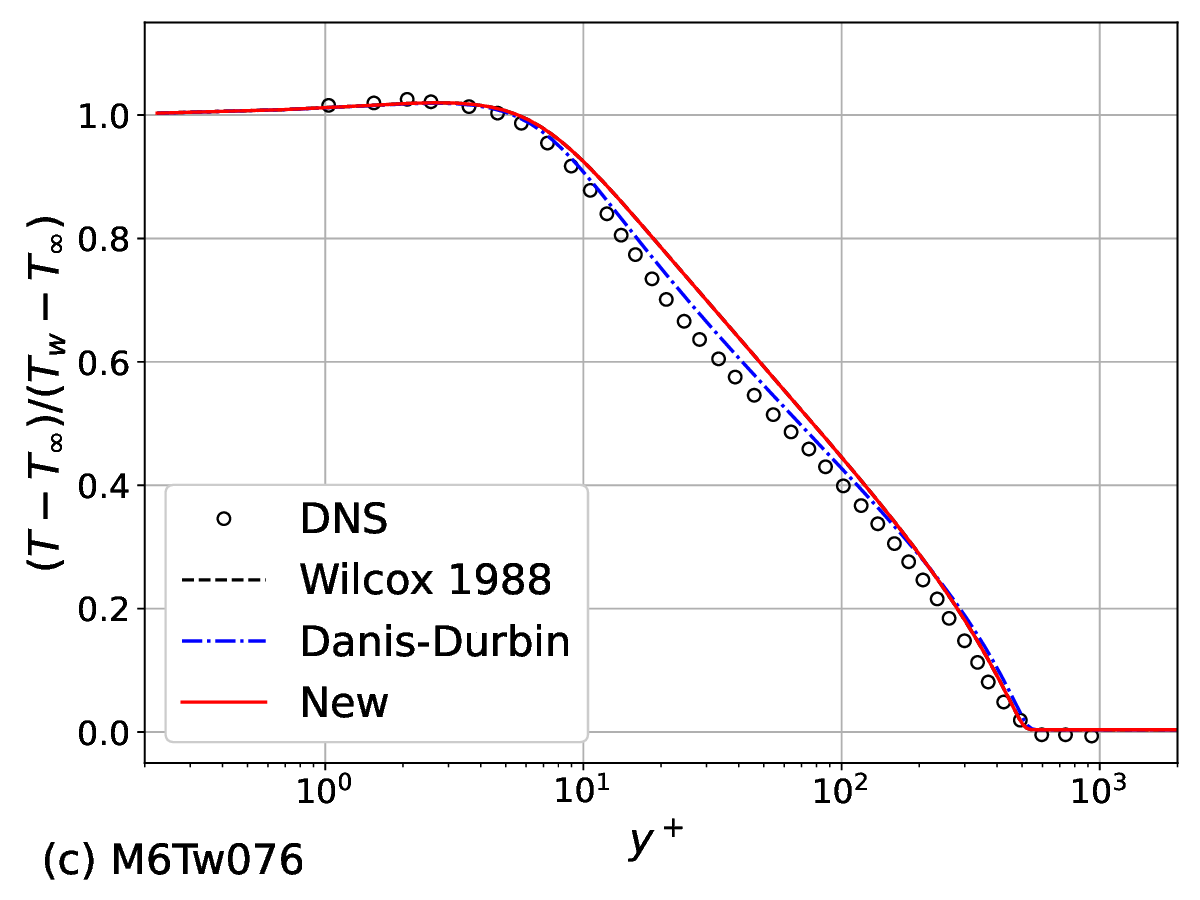}
  \end{subfigure}
  \begin{subfigure}{0.495\textwidth}
    \centering
    \includegraphics[trim={0.5cm, 0.25cm, 0.35cm, 0.25cm}, clip, width=0.9\textwidth]{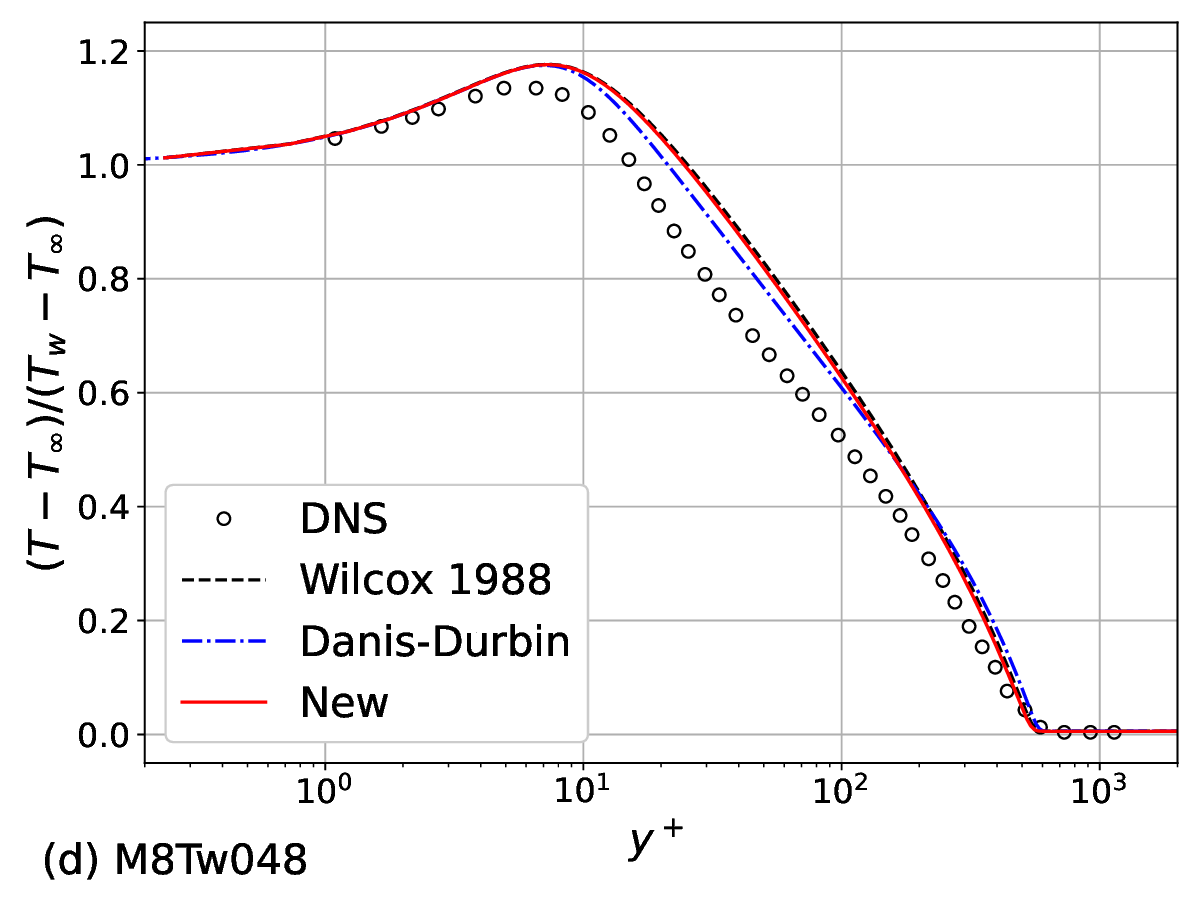}
  \end{subfigure}
  \begin{subfigure}{0.495\textwidth}
    \centering
    \includegraphics[trim={0.5cm, 0.25cm, 0.35cm, 0.25cm}, clip, width=0.9\textwidth]{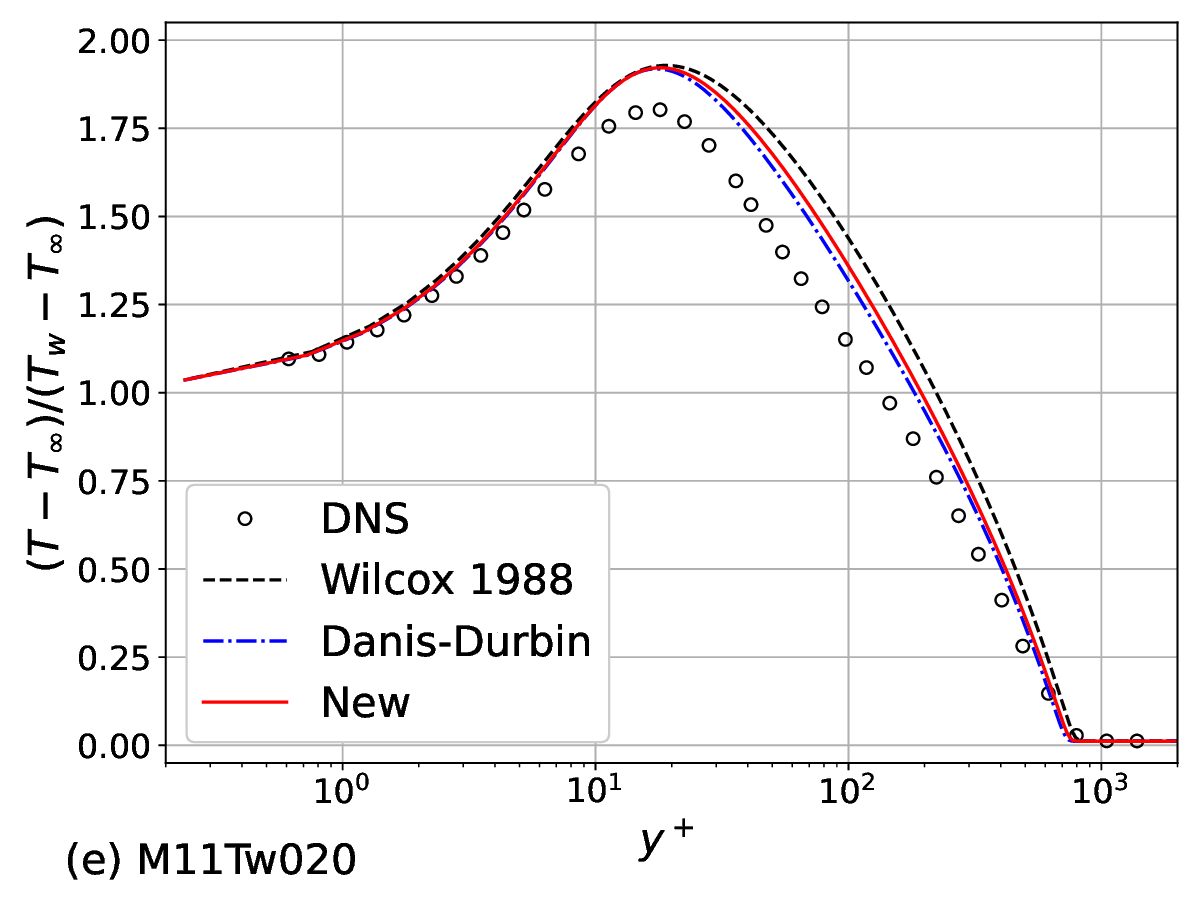}
  \end{subfigure}
  \begin{subfigure}{0.495\textwidth}
    \centering
    \includegraphics[trim={0.5cm, 0.25cm, 0.35cm, 0.25cm}, clip, width=0.9\textwidth]{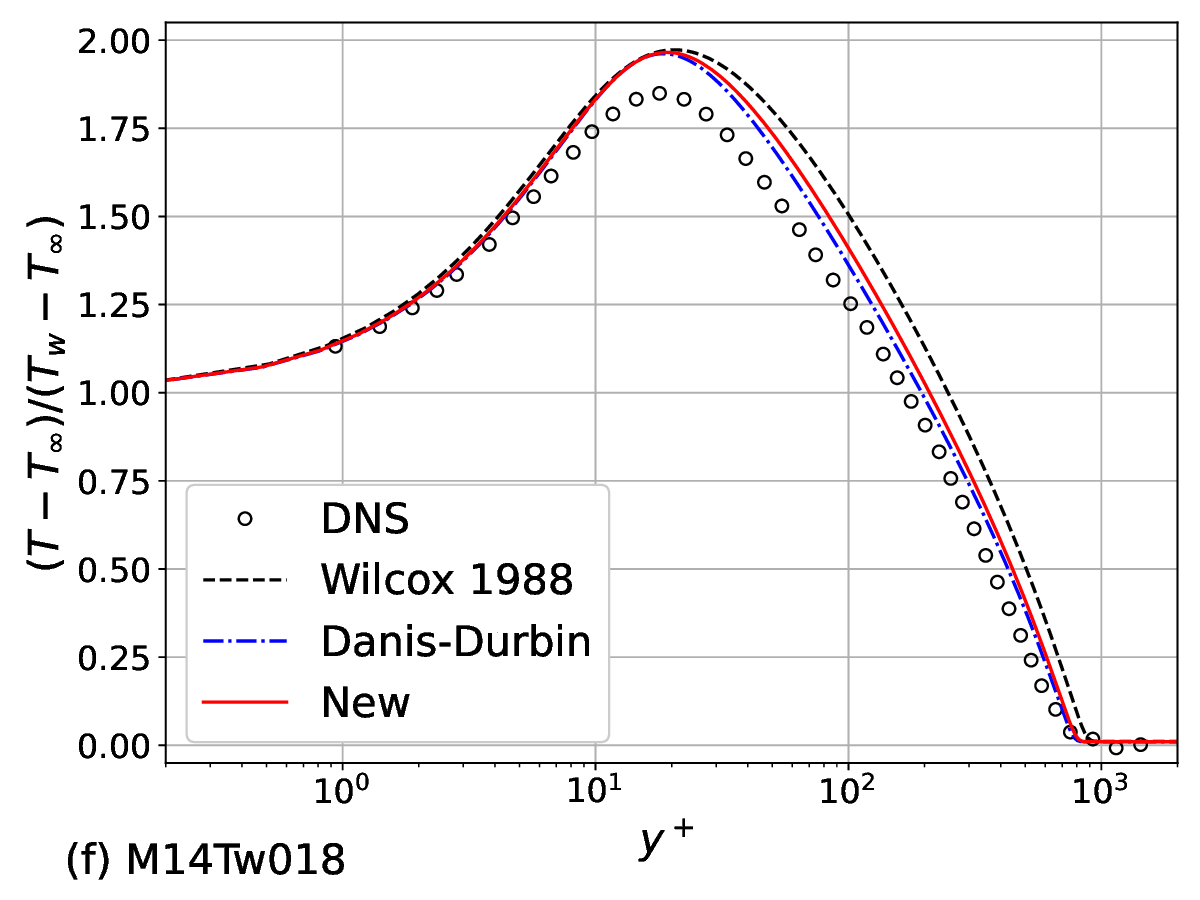}
  \end{subfigure}
  \caption{The new $k-\omega$'s predicted nondimensional temperature distribution compared to the DNS data and Wilcox's $k-\omega$ model, and the distribution of $\alpha$. The cases are (a) M2p5, (b) M6Tw025, (c) M6Tw076, (d) M8Tw048, (e) M11Tw020, and (f) M14Tw018.}
  \label{fig:TBLT}
\end{figure}

Figure \ref{fig:TBLT} shows the predicted temperature distributions for the turbulent boundary layers.
For the (a) M2p5 and (c) M6Tw076 cases, where no or very weak wall-cooling effect exists, the predicted temperature profile is identical to that of the Wilcox $k-\omega$ model, as expected.
For the (d) M8Tw048 case, where mild wall cooling invokes the empirical correction to the $k-\omega$ model, a slight improvement in the temperature distribution is also observed.
For strong wall-cooling cases, such as (b) M6Tw025, (e) M11Tw020, and (f) M14Tw018, some improvements to the temperature profiles are clearly shown, although not as significant as those of the Danis-Durbin model.
Similar to the influence on the velocity profile, the current new $k-\omega$ model's influence on the temperature profile is less aggressive than the Danis-Durbin model.
Such behavior can be easily understood, as the influence on the temperature profile is solely coming from the modification in the velocity profile and eddy viscosity.

Generally, considering the velocity and temperature distributions, the new model exhibits a more consistent behavior than the Danis-Durbin model when compared to Wilcox's $k-\omega$ model, as no influence is observed for the (a) M2p5 case, where wall cooling does not exist. 
The improvement brought by the new model becomes increasingly significant as cooling becomes stronger, \textit{e.g.} (b) M6Tw025, (e) M11Tw20, and (f) M14Tw018.

\begin{figure}
  \centering
  \begin{subfigure}{0.495\textwidth}
    \centering
    \includegraphics[trim={0.25cm, 0.25cm, 0.25cm, 0.35cm}, clip, width=0.9\textwidth]{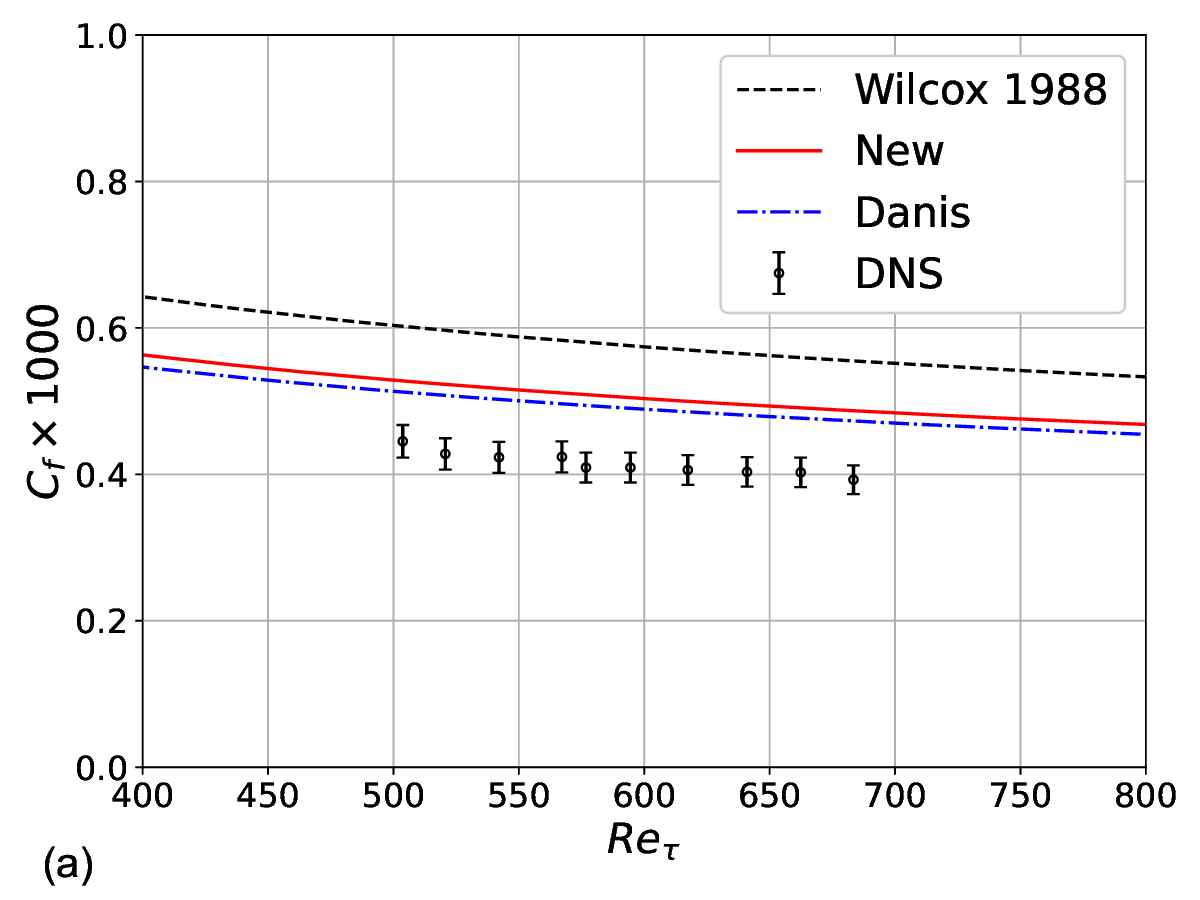}
  \end{subfigure}
  \begin{subfigure}{0.495\textwidth}
    \centering
    \includegraphics[trim={0.25cm, 0.25cm, 0.25cm, 0.35cm}, clip, width=0.9\textwidth]{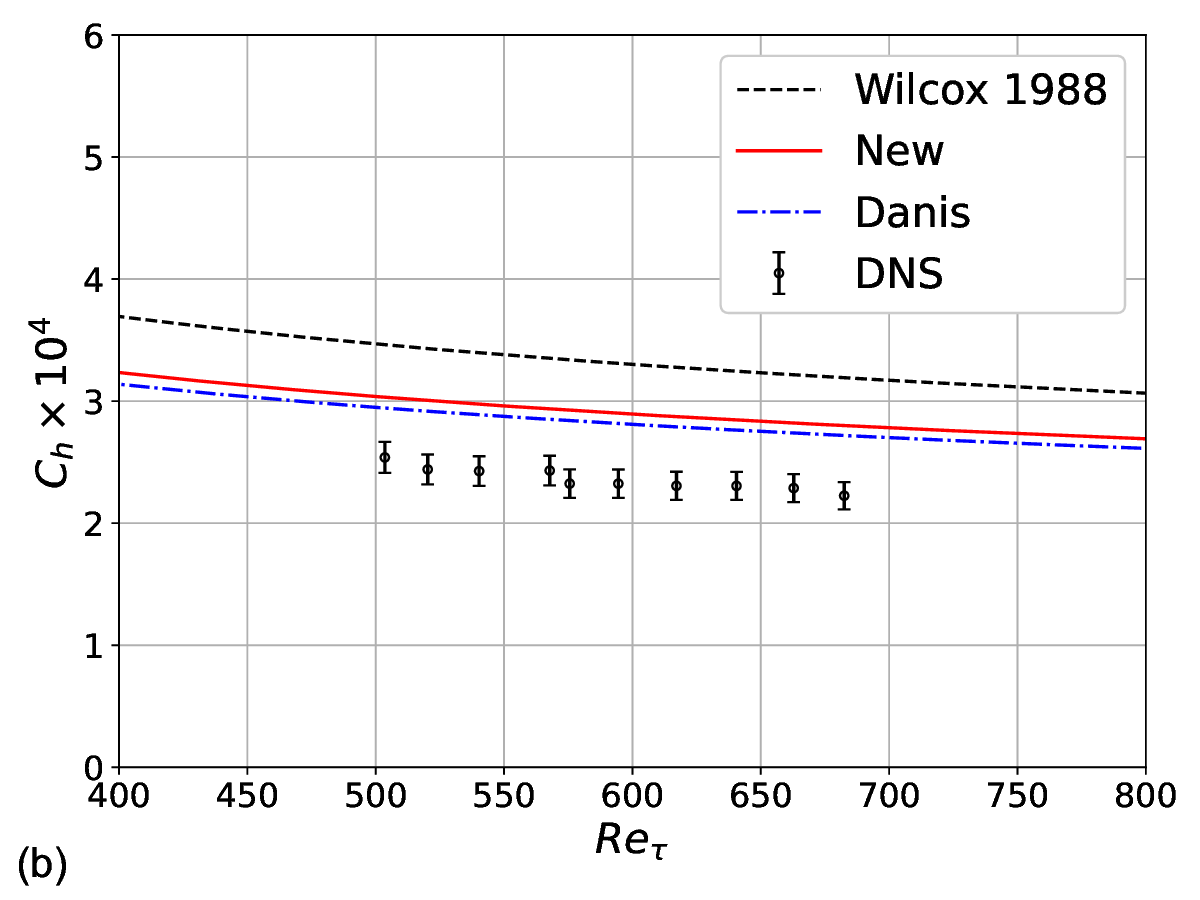}
  \end{subfigure}
  \caption{The new model's predicted skin friction and heat transfer coefficient distribution along the bottom wall, for the M14Tw018 case, compared with the Danis-Durbin model and the DNS data \cite{aiken2022assessment}.}
  \label{fig:TBLCfCh}
\end{figure}

\revised{
To examine how the new model predicts the hypersonic boundary layer evolution along the streamwise direction, surface skin friction and heat transfer coefficient distributions for the M14Tw018 case are given in Fig. \ref{fig:TBLCfCh}.
A clear improvement to the Wilcox's $k-\omega$, in terms of both the surface skin friction and the heat transfer coefficient, is observed.
Additionally, the new model's improvement, although slightly weaker, shares the same trend as the Danis and Durbin model.
The improvement is almost a direct shifting of the skin friction and heat transfer coefficient toward the DNS data, suggesting that the streamwise evolution of the new model is reasonable.
}

\section{Compressible turbulent boundary layers with varying pressure gradient}\label{sec:BL}

After testing the new model in compressible turbulent channel flows and zero-pressure-gradient boundary layers, this section assesses the model's performance under the influence of pressure gradients.

\subsection{Numerical method and flow configurations}

The numerical method used in the current simulation is the same as \S \ref{sec:numerical}.
The weak and strong favorable pressure gradient cases are from \citet{nicholson2021simulationf}, and the adverse pressure gradient case is from \citet{nicholson2021simulationa}.
Following the DNS practices, the weak and strong favorable pressure gradient cases are named M5WPG and M5SPG \cite{nicholson2021simulationf}, and the adverse pressure gradient case is named M5APG \cite{nicholson2021simulationa}.
The freestream parameters and wall boundary conditions are listed in Table \ref{tab:tbl}.
All three cases have an incoming freestream Mach number of 4.86 and a recovery temperature ratio of 0.91.
Since the recovery temperature ratio is close to unity, the wall cooling effects for the three cases are very weak.
According to the previous results on channel flows and ZPG boundary layers, the proposed new model is expected to produce predictions similar to those of Wilcox's $k-\omega$ model. 
The focus of the current tests, with pressure gradients, is to verify that the proposed modification does not bring a significant negative effect to Wilcox's $k-\omega$ model. 
The flow domain size, mesh parameters, and the number of grid points in the streamwise and spanwise directions are the same as those in the practice of \citet{danis2022compressibility}.

\subsection{Predicted results}

\begin{figure}
  \centering
  \begin{subfigure}{0.495\textwidth}
    \centering
    \includegraphics[trim={0.25cm, 0.25cm, 0.25cm, 0.25cm}, clip, width=0.9\textwidth]{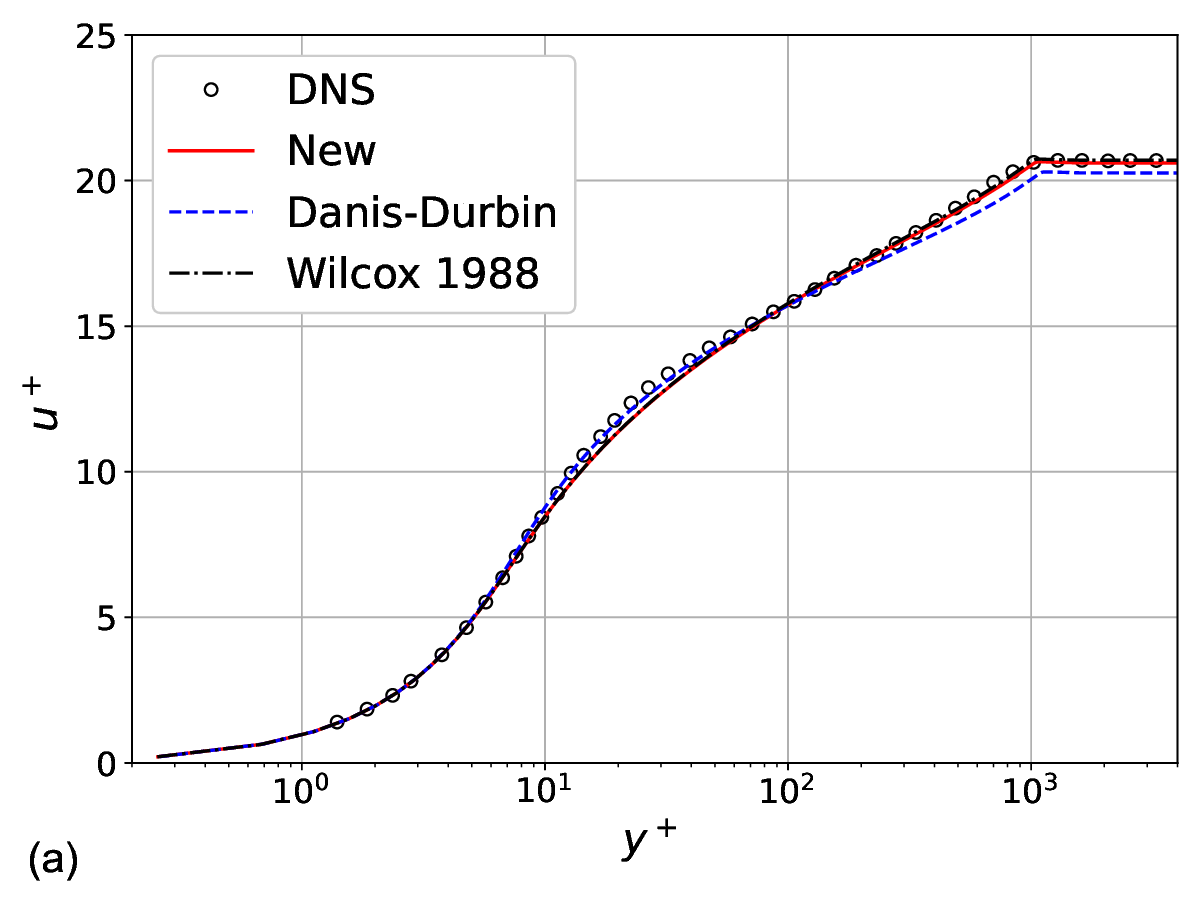}
  \end{subfigure}
  \begin{subfigure}{0.495\textwidth}
    \centering
    \includegraphics[trim={0.25cm, 0.25cm, 0.25cm, 0.25cm}, clip, width=0.9\textwidth]{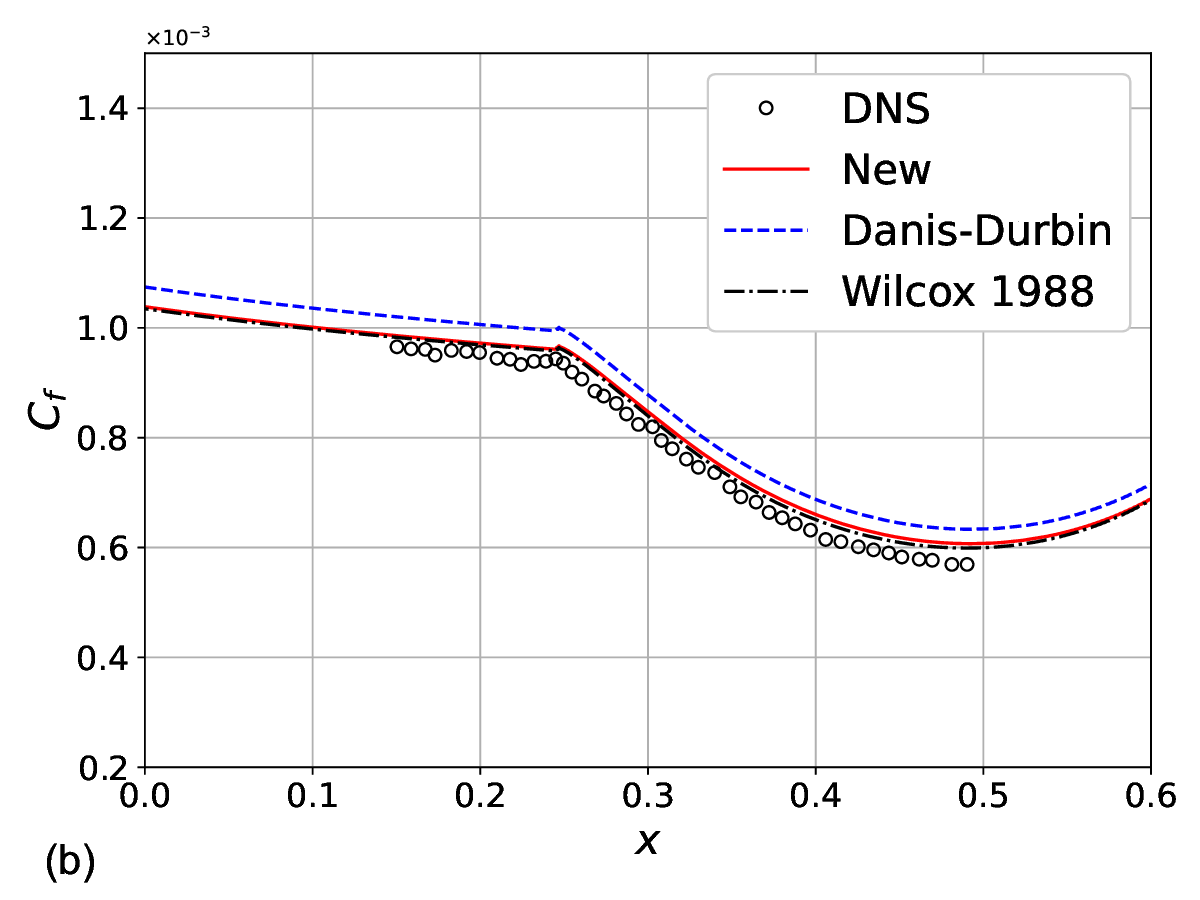}
  \end{subfigure}
  \begin{subfigure}{0.495\textwidth}
    \centering
    \includegraphics[trim={0.25cm, 0.25cm, 0.25cm, 0.25cm}, clip, width=0.9\textwidth]{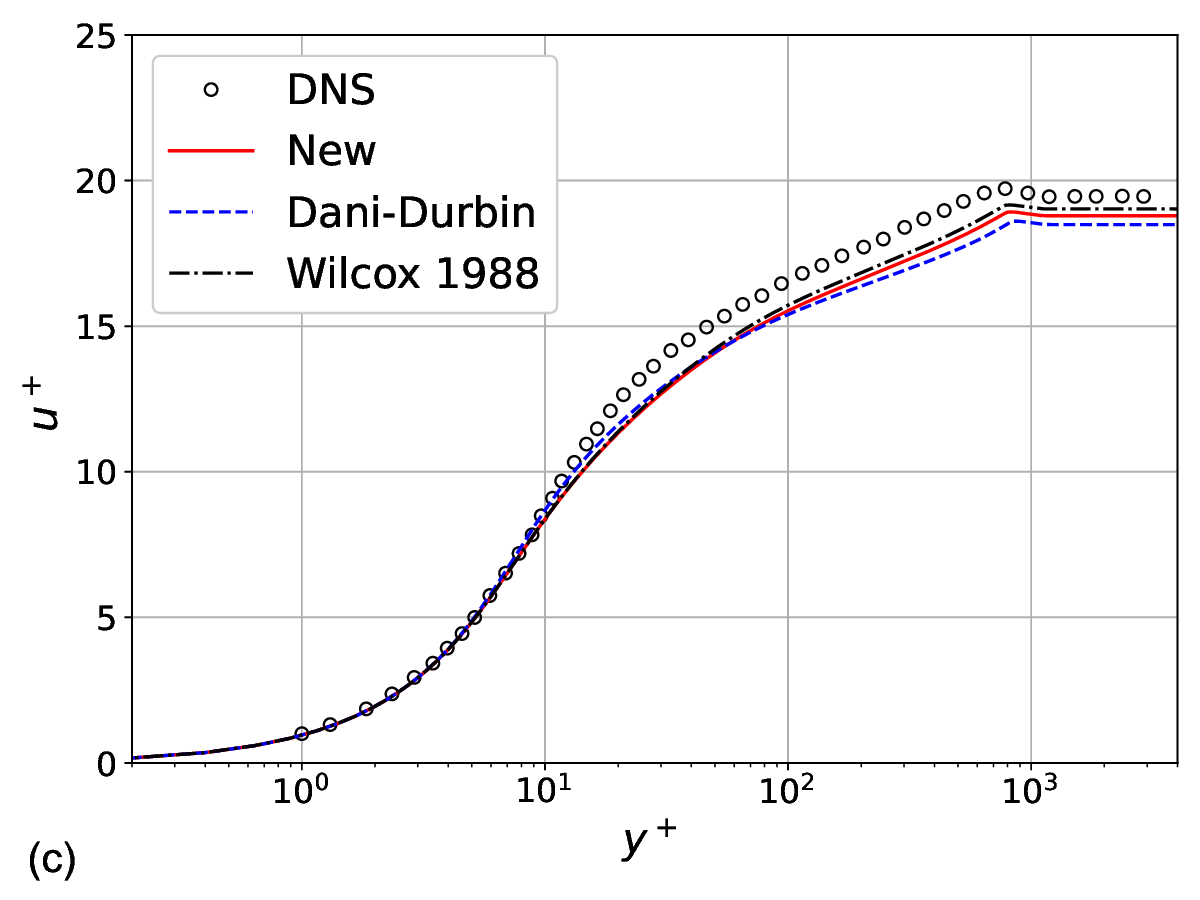}
  \end{subfigure}
  \begin{subfigure}{0.495\textwidth}
    \centering
    \includegraphics[trim={0.25cm, 0.25cm, 0.25cm, 0.25cm}, clip, width=0.9\textwidth]{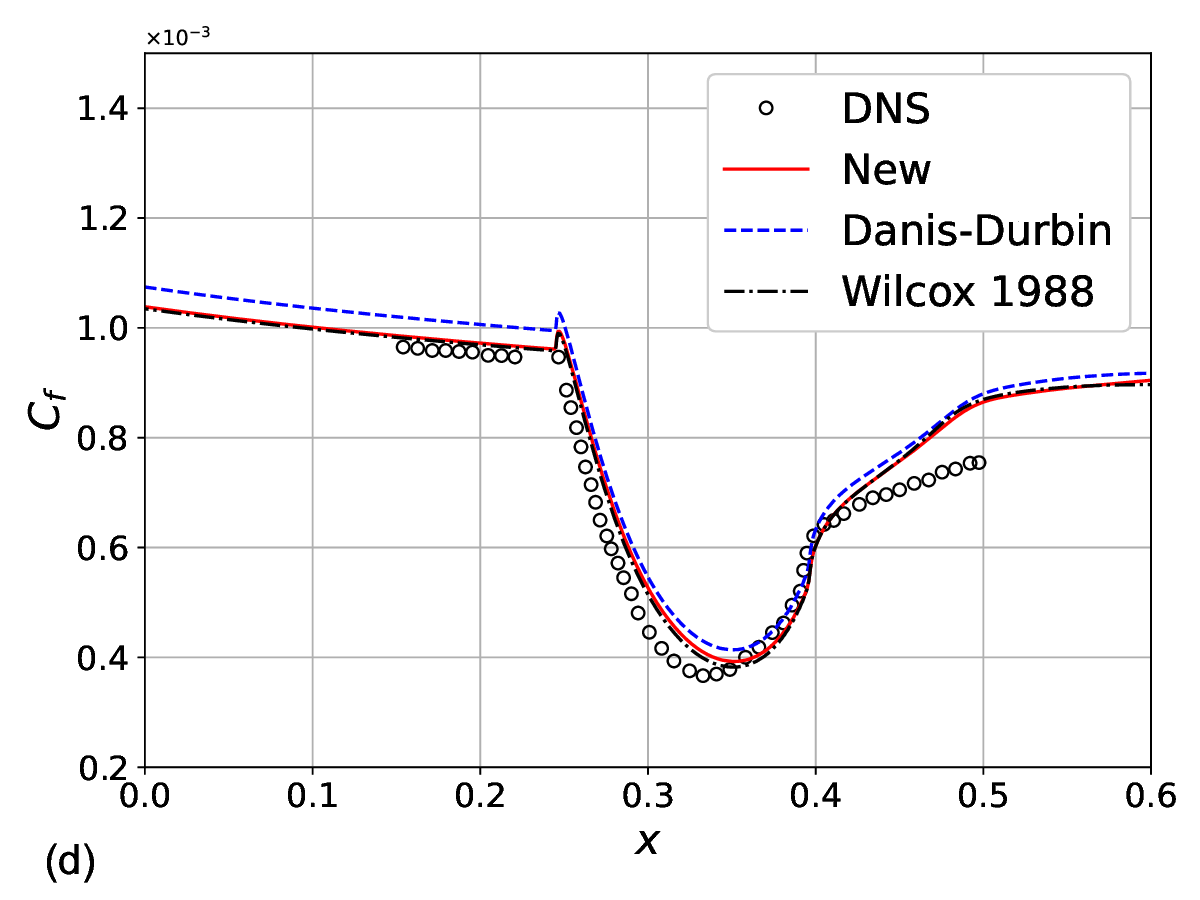}
  \end{subfigure}
  \caption{The new $k-\omega$'s predicted velocity profiles at $x=0.298$ m, and skin friction distribution under weak and strong favorable pressure gradient compared to the Wilcox 1988 $k-\omega$ model and the Danis-Durbin model. (a) velocity profile for M5WPG, (b) skin friction for M5WPG, (c) velocity profile for M5SPG, and (d) skin friction for M5SPG.}
  \label{fig:TBLFPG}
\end{figure}

Figure \ref{fig:TBLFPG} (a) and (c) show the velocity profiles for the M5WPG and M5SPG cases, while (b) and (d) show the corresponding skin friction evolution along the bottom wall.
The velocity profile is sampled at $x=0.298$, given in \citet{nicholson2021simulationf}.
For the M5WPG case, the new model predicts a velocity profile that is identical to Wilcox's 1988 $k-\omega$ model, while the Danis-Durbin model shows a slightly negative effect, as shown in Fig. \ref{fig:TBLFPG} (a).
For the M5SPG case, where the pressure gradient is stronger, the new model has a slightly negative effect on the velocity profile, while the Danis-Durbin model has the largest error.
The skin friction plots in Fig. \ref{fig:TBLFPG} (b) and (d) also confirm the observation in Fig. \ref{fig:TBLFPG} (a) and (c), that the Danis-Durbin model may not be accurate in predicting the skin friction in the flat region.
The results suggest that the new model's evolution under the influence of a favorable pressure gradient is consistent with the Wilcox 1988 $k-\omega$ model, with no apparent negative effect arising from the empirical correction in the $\omega$ equation.

\begin{figure}
  \centering
  \begin{subfigure}{0.495\textwidth}
    \centering
    \includegraphics[trim={0.25cm, 0.25cm, 0.25cm, 0.25cm}, clip, width=0.9\textwidth]{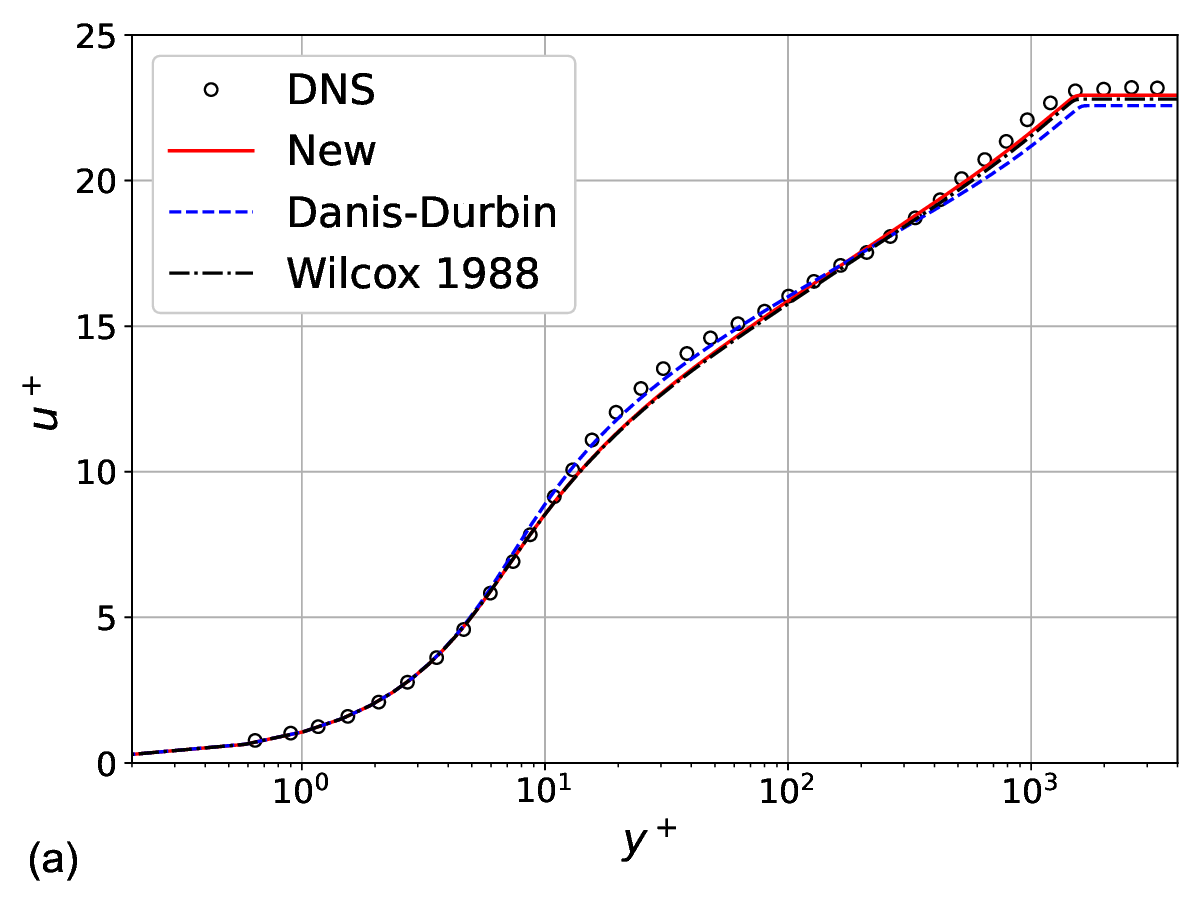}
  \end{subfigure}
  \begin{subfigure}{0.495\textwidth}
    \centering
    \includegraphics[trim={0.25cm, 0.25cm, 0.25cm, 0.25cm}, clip, width=0.9\textwidth]{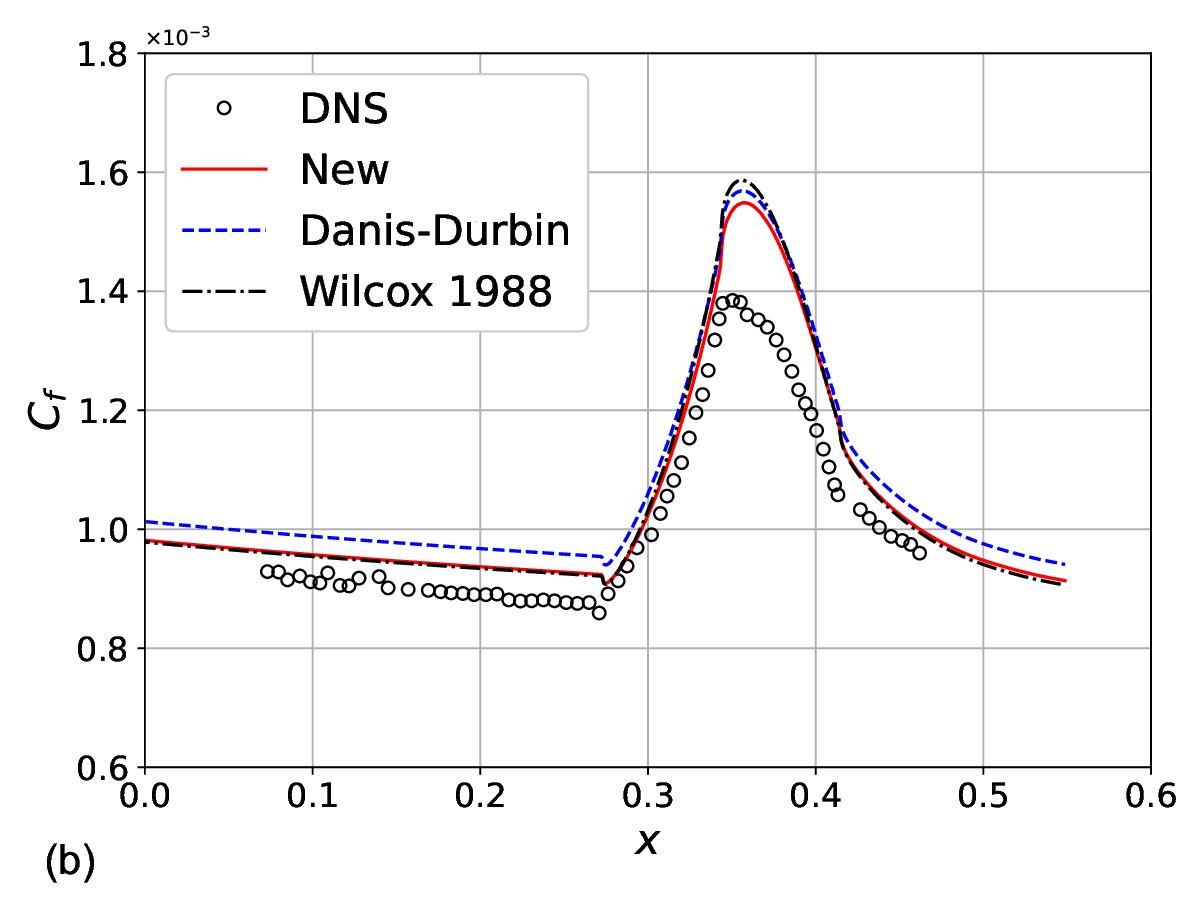}
  \end{subfigure}
    \caption{The new $k-\omega$'s predicted velocity profiles at $x=0.308$ m, and skin friction distribution under an adverse pressure gradient compared to the Wilcox 1988 $k-\omega$ model and the Danis-Durbin model. (a) velocity profile for M5APG, (b) skin friction for M5APG.}
  \label{fig:TBLAPG}
\end{figure}

The predicted result for the M5APG adverse pressure gradient case is shown in Fig. \ref{fig:TBLAPG}.
Figure \ref{fig:TBLAPG}(a) plots the velocity profile at the sampling location of $x=0.308$ m, as given in \citet{nicholson2021simulationa}.
At this location, the new $k-\omega$ model predicts an identical velocity profile compared to Wilcox's 1988 $k-\omega$ model.
The Danis-Durbin correction produces a slightly negative effect on the original $k-\omega$ model, in terms of the velocity profile.
Figure \ref{fig:TBLAPG}(b) shows the distribution of skin friction along the streamwise direction.
Along the plate, both the new model and the Danis-Durbin model produce improvement near the peak skin friction region.
The Danis-Durbin model is less accurate in terms of the skin friction in the flat plate region.
Limited by the $k-\omega$ model itself, both the new model and the Danis-Durbin model are away from the DNS data in the adverse pressure gradient region.

\section{Conclusion}\label{sec:conclusion}
This paper proposes an empirical, entirely local formula for the near-wall compressibility correction to the $k-\omega$ model, accounting for the strong wall-cooling effect in compressible wall-bounded flows.
Instead of Danis-Durbin's approach, which involves local friction Mach number, local Mach number, and local heat flux parameter, the current approach only needs a local gauge of density gradient.
The gauge is formulated using density and its gradient, together with wall distance and its gradient, via the reasoning of dimensional analysis and selection of a proper local length scale.
The gauge of local density gradient becomes an input parameter for an exponential function, which is used to suppress the source terms in the $\omega$ transport equation, following Danis-Durbin's practice of modifying the $\omega$ transport equation.
There is one model constant that comes into the exponential function, which is empirically determined.
By design, the new model recovers the original Wilcox's 1988 $k-\omega$ model \cite{wilcox1998turbulence} when the wall-cooling effect is negligible or very weak.

The new model is first assessed using a series of isothermal compressible turbulent channel flows, with a bulk Mach number no greater than 4.0 and the $Re_\tau^*$ less than 600.
The results show that the proposed new model predicts velocity profiles that are nearly identical to those of the Danis-Durbin model \cite{danis2022compressibility}.
The velocity profiles, eddy viscosity ratio, normalized turbulent kinetic energy $k$, and normalized eddy frequency $\omega$ follow certain semi-local wall scalings after the Trettel-Larsson transformation.

Furthermore, the new model is tested in a series of zero-pressure-gradient compressible turbulent boundary layers with varying wall-cooling strengths.
The proposed model exhibits a notable improvement over the Wilcox 1988 $k-\omega$ model, with performance very similar to that of the Danis-Durbin model.

At last, the new model is tested under both favorable and adverse pressure gradients, with a freestream Mach number of 4.86.
Without significant wall cooling, the new model's behavior is consistent with Wilcox's $k-\omega$ model, suggesting that the model does not bring negative effects to the $k-\omega$ model under pressure gradients.

Overall, the solution efficiency, convergence properties, and robustness of the proposed new model are similar to those of the original $k-\omega$ model.
The proposed new model only introduces an exponential function to suppress the source terms of the $\omega$ equation.
The exponential function is entirely formulated based on quantities such as density, wall distance, and their gradients.
Thus, the new model is Galilean invariant and is based entirely on local parameters.

\section{Acknowledgment}
The authors gratefully acknowledge financial support from the National Natural Science Foundation of China (No. 12472226).
The numerical computations were performed using Siyuan at the Center for High-Performance Computing, Shanghai Jiao Tong University.
SJTU Kunpeng \& Ascend Center of Excellence partially supported this work.
The author also thanks Prof. Guowei He and Prof. Paul Durbin for their support and encouragement in exploring turbulence models throughout the years.

\bibliography{sample}

\end{document}